\documentclass[reprint,prl,footinbib,floatfix,twocolumn,longbibliography,superscriptaddress]{revtex4-2}
\usepackage{times}
\usepackage{amssymb}
\usepackage{color}
\usepackage{amsmath}
\usepackage{amsbsy}
\usepackage{amsthm}
\usepackage{graphicx}
\usepackage{bbm}
\usepackage{bm}
\usepackage{epsfig}
\usepackage{xfrac}
\usepackage{xcolor}
\usepackage{enumerate}
\usepackage{multirow}
\usepackage{physics}
\usepackage[T2A,T1]{fontenc}
\usepackage{ tipa }
\usepackage[english]{babel}
\usepackage{symbols}
\usepackage{makecell}
\usepackage{appendix}
\usepackage{dsfont}
\usepackage{float}
\usepackage{pdfpages}
\usepackage{bbold}
\usepackage{placeins}
\usepackage{relsize}
\usepackage{xcolor}
\usepackage{scalerel}
\usepackage{tikz}
\usetikzlibrary{svg.path}

\DeclareMathSymbol{\shortminus}{\mathbin}{AMSa}{"39}

\definecolor{orcidlogocol}{HTML}{A6CE39}
\tikzset{
  orcidlogo/.pic={
    \fill[orcidlogocol] svg{M256,128c0,70.7-57.3,128-128,128C57.3,256,0,198.7,0,128C0,57.3,57.3,0,128,0C198.7,0,256,57.3,256,128z};
    \fill[white] svg{M86.3,186.2H70.9V79.1h15.4v48.4V186.2z}
                 svg{M108.9,79.1h41.6c39.6,0,57,28.3,57,53.6c0,27.5-21.5,53.6-56.8,53.6h-41.8V79.1z M124.3,172.4h24.5c34.9,0,42.9-26.5,42.9-39.7c0-21.5-13.7-39.7-43.7-39.7h-23.7V172.4z}
                 svg{M88.7,56.8c0,5.5-4.5,10.1-10.1,10.1c-5.6,0-10.1-4.6-10.1-10.1c0-5.6,4.5-10.1,10.1-10.1C84.2,46.7,88.7,51.3,88.7,56.8z};
  }
}

\newcommand\orcid[1]{\href{https://orcid.org/#1}{$\,$\mbox{\scalerel*{
\begin{tikzpicture}[yscale=-1,transform shape]
\pic{orcidlogo};
\end{tikzpicture}
}{|}}}}

\definecolor{myurlcolor}{rgb}{0.0,0.39,0.0}
\definecolor{myrefcolor}{rgb}{0.0,0.39,0.0}

\usepackage[colorlinks]{hyperref}
\hypersetup{unicode=true,
    bookmarksopen=false,
    breaklinks=false,
    pdfborder={0 0 0},
	bookmarksnumbered=false,
	pdfstartview={FitH},
	citecolor={myurlcolor},
	linkcolor={myrefcolor},
	urlcolor={myurlcolor}}

\usepackage{soul}

\makeatletter
\AtBeginDocument{\let\LS@rot\@undefined}
\makeatother 

\makeatletter
\def\maketitle{
\@author@finish
\title@column\titleblock@produce
\suppressfloats[t]}
\makeatother

\begin{document}

\title{Distributed Phase-Insensitive Displacement Sensing}

\author{Piotr~T.~Grochowski\orcid{0000-0002-9654-4824}}
\email{piotr.grochowski@upol.cz}
\affiliation{Department of Optics, \href{https://ror.org/04qxnmv42}{Palacký University}, 17. listopadu 1192/12, 771 46 Olomouc, Czech Republic}

\author{Matteo Fadel\orcid{0000-0003-3653-0030}}
\email{fadelm@phys.ethz.ch}
\affiliation{Department of Physics, \href{https://ror.org/05a28rw58}{ETH Zürich}, 8093 Zürich, Switzerland}

\author{Radim~Filip\orcid{0000-0003-4114-6068}}
\email{filip@optics.upol.cz}
\affiliation{Department of Optics, \href{https://ror.org/04qxnmv42}{Palacký University}, 17. listopadu 1192/12, 771 46 Olomouc, Czech Republic}

\begin{abstract}
Distributed quantum sensing leverages quantum correlations among multiple sensors to enhance the precision of parameter estimation beyond classical limits.
Most existing approaches target phase estimation and rely on a shared phase reference between the signal and the probe, yet many relevant scenarios deal with regimes where such a reference is absent, making the estimation of force or field amplitudes the main task.
We study this phase-insensitive regime for bosonic sensors that undergo identical displacements with common phases randomly varying between experimental runs.
We derive analytical bounds on the achievable precision and show that it is determined by first-order normal correlations between modes in the probe state, constrained by their average excitations.
These correlations yield a collective sensitivity enhancement over the standard quantum limit, with a gain that grows linearly in the total excitation number, revealing a distributed quantum advantage even without a global phase reference.
We identify families of multimode states with definite joint parity that saturate this limit and can be probed efficiently via local parity measurements already demonstrated or emerging in several quantum platforms.
We further demonstrate that experimentally relevant decoherence channels favor two distinct sensing strategies: splitting of a single-mode nonclassical state among the modes, which is robust to loss and heating, and separable probes, which are instead resilient to dephasing and phase jitter.
Our results are relevant to multimode continuous platforms, including trapped-ion, solid-state mechanical, optomechanical, superconducting, and photonic systems.
\end{abstract}

\maketitle

\textit{Introduction}---Quantum metrology seeks to exploit nonclassical correlations to enhance precision beyond the standard quantum limit (SQL)~\cite{Giovannetti2004,Paris2009,Degen2017}.
In continuous-variable systems, this enhancement is typically achieved through nonclassical single-mode state engineering~\cite{Duivenvoorden2017,Braun2018,Fadel2025,Rouviere2024,Pan2025,Facon2016,Burd2019,Gilmore2021,Delakouras2023,Rakhubovsky2024,Satzinger2018,Mason2019,Wollack2022,vonLupke2022,Bild2023,Youssefi2023,Millen2020,Gonzalez-Ballestero2021} or from multimode entanglement, which underlies distributed quantum sensing (DQS) protocols~\cite{Proctor2018,Zhuang2018,Eldredge2018,Guo2020,Zhang2021}.
Most DQS studies address phase estimation, assuming a well-defined phase reference shared among the modes with respect to the measured signal and relying on Gaussian resources, collective squeezing, GHZ-type entanglement, or error correction~\cite{Zhuang2018,Guo2020,Zhuang2020,Zhang2021,Malia2022,Kim2024,Hong2025,Kim2025}.
However, many sensing platforms operate in regimes where such a global reference is inaccessible or irrelevant, targeting instead the estimation of signal amplitudes.
This phase-insensitive regime is directly relevant to a broad range of systems, including trapped ions~\cite{Wolf2019,Deng2024,Valahu2025,Bond2025}, solid-state oscillators~\cite{Bild2023,Rahman2025}, optomechanical devices~\cite{Gavartin2012,Mason2019,Youssefi2023}, and superconducting resonators~\cite{Eickbusch2022,Pan2025}, and underpins applications such as Raman spectroscopy~\cite{Andrade2020}, gravitational-wave detection~\cite{Ballantini2003,LIGOO4DetectorCollaboration2023,Carney2025}, and dark-matter searches~\cite{Sikivie1983,Teufel2009,Backes2021,Brady2022,Higgins2024,Freiman2025}.

\begin{figure}[ht!]
    \includegraphics[width=0.999\linewidth]{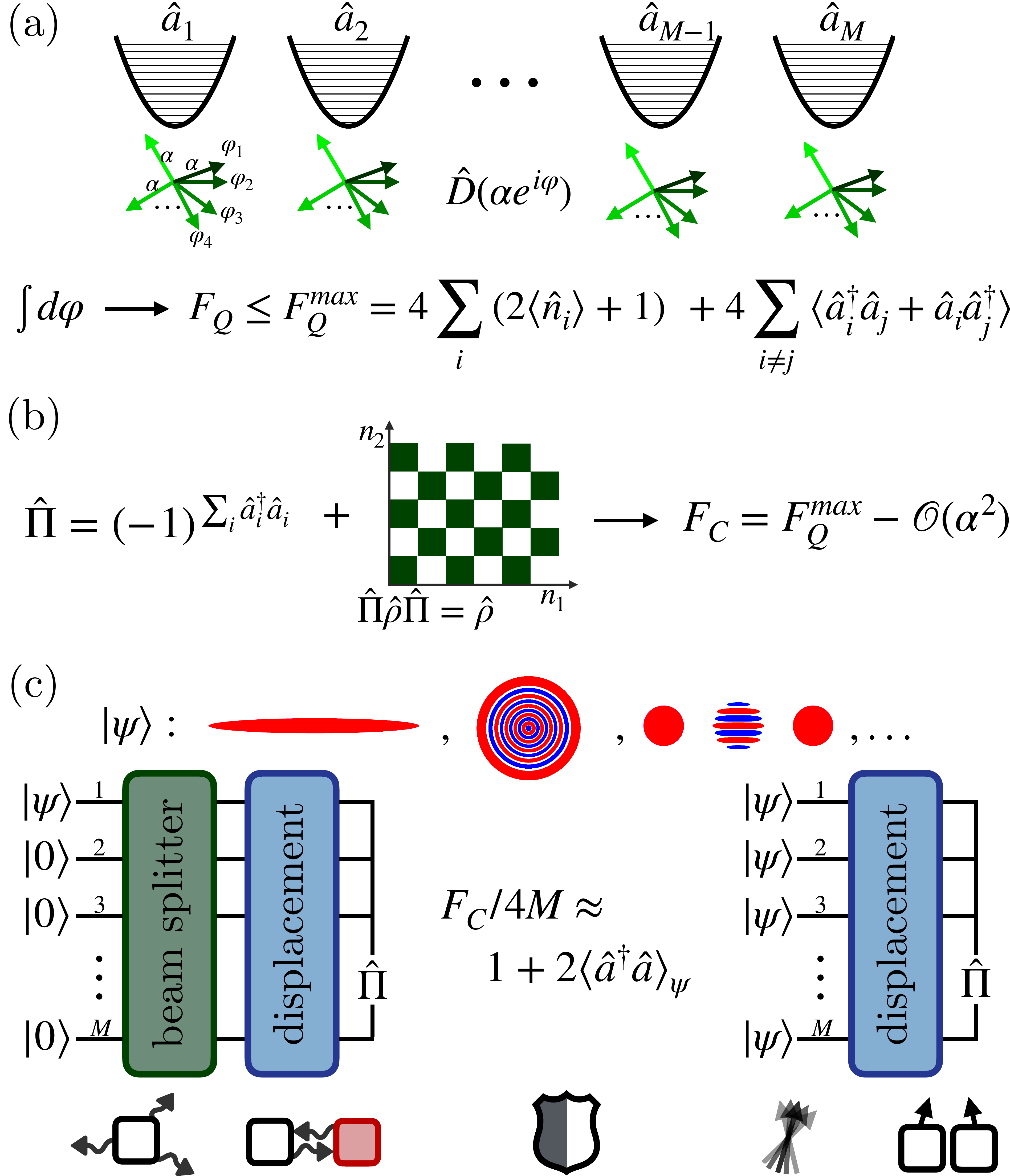}  
    \caption{\textit{Considered setup and main results.} (a) The sensing system consisting of $\ModeNumber$ bosonic modes. In the $j^{\text{th}}$ experimental shot, all the modes experience a phase-space displacement with common amplitude $\PhaseSpaceDispAmp$ and $\DispPhase_j$. The phase is not fixed between the shots and becomes fully randomized after many experimental shots~\cite{Grochowski2025}.  We show that the quantum Fisher information $\QFisher$ for estimating $\PhaseSpaceDispAmp$ is bounded by the mode occupations $\langle \NumberOperator_i\rangle$ and first-order bosonic correlations, totaling $\QFisher^{\text{max}}$. (b) Pure states with definite joint parity, $\ParityOp \DensityMatrix \ParityOp = \DensityMatrix$, giving a checkerboard structure in Fock space, can saturate this bound in the small-$\PhaseSpaceDispAmp$ limit under joint-parity measurements $\ParityOp$ (implemented via local parity, collective operations, or excitation-resolving detectors), with $\CFisher = \QFisher^{\text{max}} - \mathcal{O} (\PhaseSpaceDispAmp^2)$, where $\CFisher$ is the classical Fisher information. (c) We analyze two sensitivity-optimal measurement strategies subject to single-mode state preparation with at most $\MaxOccupation$ average excitations per mode.
In the first, a single-mode state $\ket{\WaveFunction}$ of definite parity (e.g., squeezed vacuum, Fock, or cat) is split across many modes; in the second, the same state $\ket{\WaveFunction}$ is prepared independently in each mode.
Both yield $\CFisher/\QFisher^{\mathrm{SQL}} = 1 + 2\langle \Creation \Annihilation \rangle_\WaveFunction$, with the standard quantum limit (SQL) given by $\QFisher^{\mathrm{SQL}} = 4 \ModeNumber$.
Notably, the first strategy achieves a metrological gain over the SQL that scales linearly with the total excitation number $\langle \FullNOp \rangle = \langle \sum_i \NumberOperator_i\rangle = \langle \Creation \Annihilation \rangle_\WaveFunction$.
Which strategy is optimal depends on the dominant noise: the first is more robust to excitation loss and heating, while the second better suppresses dephasing and intermode phase fluctuations.
  \label{Fig1}}
\end{figure}

Distributed force sensing in continuous-variable systems has been analyzed in the regime of fixed displacement direction, where Heisenberg scaling can be ideally achieved for most entangled probes~\cite{Kwon2022,Guo2025}, mirroring the phase-estimation result~\cite{Oszmaniec2016}.
While multimode random displacement channels, which randomize both the amplitude and phase of the displacement across modes, have been investigated~\cite{Oh2024}, channels that randomize only the displacement phase from shot to shot have so far been explored solely in single-mode settings~\cite{Wolf2019,Gorecki2022,Oh2020,Ritboon2022,Grochowski2025a}.
Here, we bridge these directions by considering $\ModeNumber$ bosonic modes subjected to identical phase-space displacements of common amplitude $\PhaseSpaceDispAmp$ and random global displacement phase that varies independently between experimental shots.
We derive analytical bounds on the quantum Fisher information (QFI) for amplitude estimation.
We show that the achievable sensitivity is set by first-order coherence between the modes, which are bounded by the mean single-mode excitation number and can be generated by splitting a separable state using beam-splitter interactions.
This yields an asymptotic Heisenberg scaling $\QFisher \sim \ModeNumber \langle \FullNOp \rangle$, which is the same as for the distributed phase-sensitive force sensing~\cite{Kwon2022}, and $\langle \FullNOp \rangle$ is the total number of excitations.

We identify families of multimode states with definite joint parity---dubbed \textit{checkerboard} states for their structured Fock-space populations---that saturate the coherence-based bound under joint-parity measurement, realized via only local parity measurements.
It constitutes a binary observable demonstrated in circuit-quantum-electrodynamical (QED)~\cite{Riste2013,Huai2024,Hinderling2024}, trapped-ion~\cite{Leibfried1996,Monroe1996,Jia2022,Jeon2025}, and photonic platforms~\cite{Lee2025}.
We further determine which checkerboard states are optimal, saturating the QFI at fixed excitation number.
Many of these optimal states are non-Gaussian, proving better under some types of decoherence and establishing a direct connection with ongoing efforts in non-Gaussian quantum metrology~\cite{Walschaers2021,Lachman2022,Hanamura2023,Rakhubovsky2024}.
Finally, we analyze robustness under realistic noise---loss, heating, dephasing, and intrashot phase jitter---and show how the advantage persists under experimental conditions, with different sensing strategies optimal in different regimes.

\textit{Sensitivity bound}---As a starting point we consider $\ModeNumber$ parallel bosonic modes with the classical commutation relations, $[\Annihilation_i,\Creation_j]=\delta_{ij}$ for $i \in \curlies{1,\dots,\ModeNumber}$.
We study a classically correlated process that displaces each of the modes with the same strength and phase in each experimental shot, $\Displacement{\PhaseSpaceDispAmp,\DispPhase} = \bigotimes_{i=1}^\ModeNumber \hat{D}_i\rounds{\PhaseSpaceDispAmp,\DispPhase} = \bigotimes_{i=1}^\ModeNumber \exp \small(\ImagUnit \PhaseSpaceDispAmp \Generator_i \small) $ with a general quadrature $\Generator_i = \small( e^{\ImagUnit \DispPhase} \Creation_i -  e^{-\ImagUnit \DispPhase} \Annihilation_i \small)/\ImagUnit$.
During each of the experimental shots, the phase $\DispPhase$ is random and drawn from the uniform distribution, such that the final phase-mixing reads
\begin{align}
    \DensityMatrix_\DispLetter =\int_{0}^{2 \pi} \frac{\dd \DispPhase}{2 \pi} \ \Displacement{\PhaseSpaceDispAmp,\DispPhase}
    \DensityMatrix \ \hat{D}^\dagger \rounds{\PhaseSpaceDispAmp,\DispPhase},
    \label{mixing}
\end{align}
where $\DensityMatrix$ is the multimode probe state.
Such a channel is a multimode generalization of a type of noisy spreading channel~\cite{Gorecki2022}.
Here, the aim is to estimate the real displacement amplitude $\PhaseSpaceDispAmp$ with maximal precision by preparing a probe, evolving it under the sensing Hamiltonian for time $\Time$, and measuring, while keeping the number of repetitions minimal.
To assess the sensitivity of such a scheme, we use the Cram\'er-Rao bound~\cite{Holevo2011}, $\Delta \tilde{\PhaseSpaceDispAmp} \geq 1/ [\sqrt{\mathcal{N}} \sqrt{\CFisher (\DensityMatrix_\PhaseSpaceDispAmp, \{\Measurement_k\})}]$,
where $\Delta \tilde{\PhaseSpaceDispAmp}$ is the root-mean-square error for the $\PhaseSpaceDispAmp$-estimator $\tilde{\PhaseSpaceDispAmp}$, $\mathcal{N}$ is the number of effective repetitions, and the classical Fisher information (FI) is 
\begin{align}
    \CFisher (\DensityMatrix_\PhaseSpaceDispAmp, \{\Measurement_k\}) = \sum\nolimits_{k} \rounds{\partial_{\DispLetter}  \Prob_k^{\DispLetter}}^2 / \Prob_k^{\DispLetter},
\end{align}
where $\Prob_k^{\DispLetter} = \Tr(\Measurement_k \DensityMatrix_\PhaseSpaceDispAmp)$ and $\{\Measurement_k\}$ is specific choice of the measurement.
As $\mathcal{N}$ is often limited or variable due to technical and resource constraints~\cite{Demkowicz-Dobrzanski2015,Hradil2019}, we focus on optimizing the asymptotic single-shot FI.
The FI optimized over all possible measurements is the QFI that reads
\begin{align}
    \QFisher \rounds{\DensityMatrix_\PhaseSpaceDispAmp} = \text{max}_{\{\Measurement_k\}} \CFisher (\DensityMatrix_\PhaseSpaceDispAmp, \{\Measurement_k\}) = \Tr \small(\DensityMatrix_\DispLetter \SymmDer \small),
\end{align}
where the symmetric logarithmic derivative $\SymmDer$ is given through $\dd \DensityMatrix_\DispLetter / \dd \PhaseSpaceDispAmp =  (\SymmDer \DensityMatrix_\DispLetter + \DensityMatrix_\DispLetter \SymmDer )/2$.
From the convexity of QFI and linearity of~\eqref{mixing}, the optimal state for estimating $\PhaseSpaceDispAmp$ is pure.
Let us denote $\ket{\WaveFunction_\PhaseSpaceDispAmp^\DispPhase} = \Displacement{\PhaseSpaceDispAmp,\DispPhase} \ket{\WaveFunction}$.
Then, from convexity,
\begin{align}
    \QFisher \rounds{\DensityMatrix_\DispLetter} \leq  \int_{0}^{2 \pi} \frac{\dd \DispPhase}{2 \pi} \QFisher \rounds{\ket{\WaveFunction_\PhaseSpaceDispAmp^\DispPhase}} \quad\text{for all $\alpha$}.
    \label{average}
\end{align}
We can explicitly write the $\QFisher \rounds{\ket{\WaveFunction_\PhaseSpaceDispAmp^\DispPhase}} $,
\begin{align}
    &\QFisher \rounds{\ket{\WaveFunction_\PhaseSpaceDispAmp^\DispPhase}}/4  =  \Delta^2_{\ket{\WaveFunction_\PhaseSpaceDispAmp^\DispPhase}} \sum_i \Generator_i  =   \Delta^2_{\ket{\WaveFunction}} \sum_i \Generator_i \leq  \langle (\sum_i \Generator_i)^2 \rangle \nonumber \\
    & =-   e^{\ImagUnit  2 \DispPhase}  \langle(\sum_i \Creation_i )^2\rangle - e^{-\ImagUnit  2 \DispPhase}  \langle(\sum_i \Annihilation_i )^2\rangle + \langle \sum_{i,j} \Creation_i \Annihilation_j + \Annihilation_i \Creation_j  \rangle.
    \label{before_mixing}
\end{align}
Performing the integral in~\eqref{average}, the $\DispPhase$-dependent terms vanish,
\begin{align}
    \QFisher \rounds{\DensityMatrix_\DispLetter} &\leq 4 \sum_i (2 \langle{\Creation_i \Annihilation_i} \rangle +1)  + 4 \sum_{i \neq j} \langle{\Creation_i \Annihilation_j + \Creation_j \Annihilation_i }\rangle \label{full_bound} \\
    &\leq  4 \sum_i (2 \langle{\Creation_i \Annihilation_i} \rangle +1) + 8  \sum_{i \neq j} \sqrt{ {\langle{\Creation_i \Annihilation_i}\rangle\langle{\Creation_j \Annihilation_j}\rangle}}
    \label{full_bound2} \\
    &\leq  4 \sum_i (2 \langle{\Creation_i \Annihilation_i} \rangle +1) + 8  (\ModeNumber-1) \sum_i  \langle{\Creation_i \Annihilation_i} \rangle \\
    &= 4\ModeNumber + 8 \ModeNumber \sum_i \langle{\Creation_i \Annihilation_i} \rangle ,
    \label{full_bound3}
\end{align}
where the inequality~\eqref{full_bound2} comes from the Cauchy-Schwartz inequality, $|\langle{\Creation_i \Annihilation_j }\rangle|^2 \leq |\langle{\Creation_i \Annihilation_i}\rangle| |\langle{\Creation_j \Annihilation_j}\rangle|$, while~\eqref{full_bound3} from $\sum_{i \neq j} {{\langle{\Creation_i \Annihilation_i}\rangle^{1/2}\langle{\Creation_j \Annihilation_j}\rangle^{1/2}}} = (\sum_i {\langle{\Creation_i \Annihilation_i}\rangle^{1/2}})^2 - \sum_i \langle{\Creation_i \Annihilation_i}\rangle \leq \ModeNumber \sum_i \langle{\Creation_i \Annihilation_i}\rangle   - \sum_i \langle{\Creation_i \Annihilation_i}\rangle  $.
The above derivation follows the line of thought proposed in~\cite{Gorecki2022}.
In the case we consider, the SQL can be defined as a QFI maximized over multimode classical state, $\DensityMatrix_\text{cl} = \int \dd \pmb{\CohDis} P(\pmb{\CohDis}) \bigotimes_{i=1}^{\ModeNumber} \ket{\CohDis_i}\bra{\CohDis_i}$, where $\ket{\CohDis_i}$ is a coherent state and $P(\pmb{\CohDis})>0$.
Then, due to convexity, we have $\QFisher(\DensityMatrix_\text{cl})\leq \int \dd \pmb{\CohDis} P(\pmb{\CohDis}) \QFisher(\bigotimes_{i=1}^{\ModeNumber} \ket{\CohDis_i}\bra{\CohDis_i})=4 \ModeNumber = \QFisher^{\text{SQL}}$, where the penultimate equality follows from the single-mode result~\cite{Wolf2019,Oh2020,Ritboon2022, Gorecki2022,Grochowski2025}.
Then, we can rewrite~\eqref{full_bound3} as
\begin{align}
    \QFisher \rounds{\DensityMatrix_\DispLetter}/\QFisher^{\text{SQL}} \leq 2 \langle \FullNOp \rangle+1,
     \label{asymbound}
\end{align}
where we introduced the total excitation number operator $\FullNOp = \sum_i \Creation_i \Annihilation_i$.
This result is consistent with the previously derived bound for a single-mode system, $\ModeNumber = 1$, $\QFisher \rounds{\DensityMatrix_\DispLetter} \leq  4(2 \langle{\Creation \Annihilation} \rangle +1)$~\cite{Wolf2019,Oh2020,Gorecki2022,Grochowski2025}.
Eq.~\eqref{full_bound} allows us to define two additive parts---self contribution, $\SMQFisher\rounds{\DensityMatrix_\DispLetter} = 4 \sum_i (2 \langle{\Creation_i \Annihilation_i} \rangle +1)$, and a cross-mode enhancement due to bosonic correlations, $\MMQFisher = 4 \sum_{i \neq j} \langle{\Creation_i \Annihilation_j + \Creation_j \Annihilation_i }\rangle$.
We provide a comparison with the phase-sensitive case in App.~\hyperref[appA]{A}.

\textit{Saturating the bound~\eqref{full_bound}}---Let us now focus on finding probe states and measurements that saturate~\eqref{full_bound} for small $\PhaseSpaceDispAmp>0$.
It suffices to consider the joint parity measurement,
\begin{align}
    \Measurement^{\Parity}_{\pm} = [1 \pm \ParityOp]/2 =   [1 \pm (-1)^{\sum_i \Creation_i \Annihilation_i}]/2
\end{align}
which is a binary observable local in the physical modes, therefore scalable, that has been realized experimentally in various platforms~\cite {Huai2024,Hinderling2024,Jia2022,Jeon2025,Lee2025}.
The classical FI associated with such a binary measurement reads
\begin{align}
    \CFisher (\Measurement^{\Parity},\DensityMatrix_\PhaseSpaceDispAmp) = (\partial_\PhaseSpaceDispAmp \langle \ParityOp \rangle_{\PhaseSpaceDispAmp})^2/[1 - \langle \ParityOp \rangle_{\PhaseSpaceDispAmp}^2],
\end{align}
where $\langle \ParityOp \rangle_{\PhaseSpaceDispAmp} = \Tr \small(\DensityMatrix_\DispLetter \ParityOp \small)$.
Notably, if $|\langle \ParityOp \rangle_{\PhaseSpaceDispAmp}| = 1 - \SomeConstant \PhaseSpaceDispAmp^2 + \mathcal{O}(\PhaseSpaceDispAmp^3)$, the FI does not vanish for small $\PhaseSpaceDispAmp >0$, $\CFisher (\Measurement^{\Parity},\DensityMatrix_\PhaseSpaceDispAmp) = 2 \SomeConstant +  \mathcal{O}(\PhaseSpaceDispAmp)$.
The condition $|\langle \ParityOp \rangle_{\PhaseSpaceDispAmp=0}|=1$, necessary for such a case, is fulfilled by states $\DensityMatrix$ with definite parity, $\ParityOp \DensityMatrix  = \ParityValue \DensityMatrix$ with $\ParityValue = \pm 1$ (we will use the moniker \textit{checkerboard states} as their excitation distribution in the Fock basis forms a checkerboard).
Let us introduce the common (bright) mode, $\CommAnnihilation =  \sum_{i=j}^{\ModeNumber} \Annihilation_j / \sqrt{\ModeNumber}$. 
Then, the full multimode displacement operator can be written as a single-mode displacement in the common mode, $\Displacement{\PhaseSpaceDispAmp,\DispPhase} = \hat{D}_\BLetter(\sqrt{\ModeNumber}\PhaseSpaceDispAmp,\DispPhase)$.
Utilizing identity $\hat{D}^\dagger (\CohDis) \ParityOp \hat{D} (\CohDis) = \hat{D} (-2\CohDis) \ParityOp $, we can write
\begin{align}
    \langle \ParityOp \rangle_{\PhaseSpaceDispAmp} = \ParityValue \int_0^{2 \pi} \frac{\dd \DispPhase}{2 \pi} \Characteristic_\BLetter (-2 \sqrt{\ModeNumber} \PhaseSpaceDispAmp e^{\ImagUnit \DispPhase}),
\end{align}
where we have introduced a characteristic function of the common mode, $\Characteristic_\BLetter (\CohDis) = \Tr \small[ \DensityMatrix \hat{D}_\BLetter (\CohDis) \small]$.
Rewriting it as a $\Characteristic_\BLetter (\CohDis) = e^{-|\CohDis|^2/2} \langle e^{\CohDis \CommCreation} e^{-\CohDis^* \CommAnnihilation} \rangle$ with Baker-Campbell-Hausdorff formula and expanding in small $\PhaseSpaceDispAmp$, we can write
\begin{align}
    \CFisher (&\Measurement^{\Parity},\DensityMatrix_\PhaseSpaceDispAmp) / 4 \ModeNumber =  1 + 2 \langle \NumberOperator_\BLetter \rangle \nonumber \\
    &- 2 \ModeNumber \PhaseSpaceDispAmp^2 (1 - 2  \langle \NumberOperator_\BLetter \rangle^2 + \langle \NumberOperator_\BLetter \rangle + 3 \langle \NumberOperator_\BLetter^2 \rangle) + \mathcal{O}(\PhaseSpaceDispAmp^4),
    \label{withcorrection}
\end{align}
where $\NumberOperator_\BLetter = \CommCreation\CommAnnihilation$. 
One can easily identify that the $\PhaseSpaceDispAmp$-independent term is 
\begin{align}
    4 \ModeNumber (1 + 2 \langle \NumberOperator_\BLetter \rangle \nonumber) = 4 \sum_i (2 \langle{\Creation_i \Annihilation_i} \rangle +1)  + 4 \sum_{i \neq j} \langle{\Creation_i \Annihilation_j + \Creation_j \Annihilation_i }\rangle,
\end{align}
meaning~\eqref{full_bound} is saturated for checkerboard states and small $\PhaseSpaceDispAmp$,
\begin{align}
    \CFisher \rounds{\DensityMatrix_\DispLetter} = \QFisher \rounds{\DensityMatrix_\DispLetter} - \mathcal{O}(\PhaseSpaceDispAmp^2),
    \label{satur}
\end{align}
where we have started to drop the explicit dependence of $\CFisher$ on the measurement for the simplicity of the notation. 
The same result~\eqref{satur} can be derived via explicit evaluation in the Fock basis and for a projective measurement on multimode Fock states, both of which we provide in Supplemental Material (SM)~\cite{SuppMat5}.
While the leading term depends only on the number of excitations in the common mode, the correction $\mathcal{O}(\PhaseSpaceDispAmp^2)$ for finite $\PhaseSpaceDispAmp$ is generally decreased for common-mode-number-squeezed states, i.e., for which $\langle \NumberOperator_\BLetter^2 \rangle$ at a given $\langle \NumberOperator_\BLetter \rangle$ is lower than for Gaussian states.
This suggests that non-Gaussian states become preferable to the Gaussian ones for measurement at finite $\PhaseSpaceDispAmp$, consistently with an analogous conclusion in a single-mode case~\cite{Grochowski2025}.

\textit{Optimal states}---Let us now analyze which checkerboard states are optimal.
Clearly, accumulating all the excitations in the common mode, $\ket{\WaveFunction}_{\text{del}} = \ket{\WaveFunctionB}_{\BLetter} \bigotimes_{i=2}^{M} \ket{0}_{\BLetter_i}$, where $\CommAnnihilation_i$ are associated to orthogonal family of modes with $\CommAnnihilation_1 = \CommAnnihilation$, will saturate~\eqref{asymbound} in the leading order of $\PhaseSpaceDispAmp$ as then $\langle\NumberOperator_\BLetter\rangle = \langle\FullNOp\rangle$. 
The choices for $\ket{\WaveFunctionB}$ need to have a definite parity, including squeezed vacuum, Fock states, cat states~\cite{Zurek2001}, grid states~\cite{Gottesman2001a}, number-phase states~\cite{Grimsmo2020}, $\ket{mn}$ states~\cite{Kovalenko2025}, generalized squeezed states~\cite{Bazavan2024}, and others.
The notable non-Gaussian example includes a multimode cat state, $\ket{\WaveFunction}_\text{cat}   \sim \ket{\CohDis}^{\otimes \ModeNumber} + \ket{-\CohDis}^{\otimes \ModeNumber}$, where $\Annihilation_i \ket{\CohDis}_i = \CohDis \ket{\CohDis}_i$ and $\forall_{i,j} \ \  {}_\text{cat}\bra{\WaveFunction} \Creation_i \Annihilation_j \ket{\WaveFunction}_\text{cat} =  |\CohDis|^2 $.
Such states can be deterministically and efficiently realized via methods such as coupling to ancilla qubits, dissipation engineering~\cite{Zapletal2022}, or by beam-splitter-based distribution of coherence~\cite{Toyoda2015,Wang2016,Gao2018a,Lu2023,Qiao2023,vonLupke2024}.
Experimentally, they have been prepared in circuit QED~\cite{Wang2016} and trapped ions~\cite{Jeon2024}.
It is worth noting that multimode generalization of number-squeezed cat states~\cite{deMatosFilho1996,Dodonov2002,Rojkov2024,Rousseau2025,Grochowski2025} might prove better suited for the considered task at finite $\PhaseSpaceDispAmp$ [cf. Eq.~\eqref{withcorrection}].

However, since the generation of such all-in-common-mode states may in some cases require control over collective modes, we restrict the considered control to local single-mode preparation and passive (linear) mode mixing.
Such control can yield \textit{passively separable} states~\cite{Walschaers2017}, namely, states that become separable in some basis of the considered modes. 
Without the loss of generality, we may assume that the state is separable in the original physical modes, and then apply a passive unitary, which can be realized via a beam-splitter cascade~\cite{Clements2016}, such that $\ket{\WaveFunction} = \BogoUnitary \bigotimes_{i=1}^{M} \ket{\WaveFunctionB_i}_{i}$.
Then, after some straightforward algebra, one can rewrite $\CFisher \rounds{\DensityMatrix_\DispLetter} = 8 \TElement^{\dagger} \CoherenceMatrix \TElement + 4 \ModeNumber$, where the coherence matrix reads $\CoherenceMatrix_{ij} = \langle \Creation_i \Annihilation_j \rangle$, $\TElement = \BogoUnitaryElement 1_\ModeNumber$ is an auxiliary vector, and $1_\ModeNumber$ is a vector of ones.
The requirement for the definite joint parity implies $\forall_i\langle \Annihilation_i\rangle =0$ and then $\CoherenceMatrix = \text{diag}(\langle \Creation_1 \Annihilation_1\rangle,\dots,\langle \Creation_\ModeNumber \Annihilation_\ModeNumber\rangle)$.
Let us now consider a physically relevant scenario, where the state preparation is limited by a local single-mode control that has a maximum average occupancy, $\forall_i\langle \Creation_i \Annihilation_i\rangle \leq \MaxOccupation$.
Then, $\TElement^{\dagger} \CoherenceMatrix \TElement = \sum_i |\TElement_i|^2 \langle \Creation_i \Annihilation_i\rangle \leq \sum_i |\TElement_i|^2 \MaxOccupation = \ModeNumber \MaxOccupation$, and
\begin{align}
\CFisher \rounds{\DensityMatrix_\DispLetter} \leq 4 \ModeNumber (2 \MaxOccupation + 1).
\label{bound_sep}
\end{align}
Notably, saturation of this inequality does not guarantee saturation of the QFI, as the maximal allowed number of excitations in the system is $\ModeNumber \MaxOccupation > \MaxOccupation$.

Let us then consider two alternative approaches that saturate~\eqref{bound_sep}.
The first, \textit{correlated}, one involves preparation of the first mode in $\ket{\WaveFunctionB'}_{a_1}$, leaving the rest empty and passing all the modes through the balanced beam-splitter cascade~\cite{Reck1994}, given, e.g., by discrete Fourier transform form, $(\BogoUnitaryElement)_{ij} = e^{2 \pi \ImagUnit (i-1) (j-1)/\ModeNumber} /\sqrt{\ModeNumber}$, producing state with excitations only in the common mode, $\ket{\WaveFunction}_\text{bs} = \BogoUnitary \ket{\WaveFunctionB'}_{a_1}\bigotimes_{j=2}^{M} \ket{0}_{a_j} = \ket{\WaveFunctionB'}_{\BLetter} \bigotimes_{i=2}^{M} \ket{0}_{\BLetter_i}$.
The latter equality comes directly from $\BogoUnitary^\dagger \Creation_1 \BogoUnitary = \CommCreation$, implying $\langle \NumberOperator_\BLetter \rangle = \MaxOccupation$ and $\CFisher \rounds{\DensityMatrix_\DispLetter} = 4 \ModeNumber (2 \MaxOccupation + 1)$.
Such a state saturates the inequality~\eqref{asymbound} as $\langle \FullNOp \rangle = \MaxOccupation$.
The second, \textit{separable}, strategy involves preparing the state $\ket{\WaveFunctionB''}$ in every single mode, $\ket{\WaveFunction}_{\text{sep}} =  \bigotimes_{i=1}^{M}\ket{\WaveFunctionB''}_{a_i}$ and explicitly saturates~\eqref{bound_sep} without any passive mode mixing.
In this case,~\eqref{asymbound} is {\em not saturated} as $\langle \FullNOp \rangle = \ModeNumber \MaxOccupation$.
Despite the same FI, the main difference comes in the distribution of the sensitivity between self and cross contributions.
For the correlated case, the cross contribution dominates as $\SMCFisher(\DensityMatrix_\PhaseSpaceDispAmp^{\text{bs}}) = 8 \MaxOccupation + 4 \ModeNumber$ and $\MMCFisher(\DensityMatrix_\PhaseSpaceDispAmp^{\text{bs}}) = 8 (\ModeNumber-1) \MaxOccupation $, while for the separable one there is only a self contribution, $\CFisher (\DensityMatrix_\DispLetter^{\text{sep}}) = \SMCFisher(\DensityMatrix_\PhaseSpaceDispAmp^{\text{sep}}) $.
In the distributed case, the metrological gain comes majorly from the correlations between the modes, while for the other case, it is only due to single-mode enhancement.
These contributions experience different decoherence susceptibilities, as shown in the next paragraph.
A comparison of these strategies and homodyne-based Gaussian schemes in an exemplary two-mode case is presented in the SM~\cite{SuppMat5}.

\textit{Decoherence}---Imperfections inevitably lead to a decrease in the sensitivity; however, different states can have different noise susceptibility.
Both imperfect state preparation and decoherence during the interrogation can lead to reduced FI.
To model how the FI is affected, we will analyze several common sources of noise in the probe state: excitation loss, mechanical heating, mechanical dephasing, and intermode phase jitter during a single experimental shot. 
The starting point for the first three is the master equation for the evolution of the bosonic modes coupled to the bath with thermal occupation $\ThermalOccupation$ in time $\Time$~\cite{Breuer2007},
\begin{align}
    \partial_{\Time} \DensityMatrix (\Time) &=  \sum_{i,j} \frac{1}{2} \squares{2 \Coll_{ij} \DensityMatrix(\Time) \Coll_{ij}^{\dagger} - \DensityMatrix(\Time ) \Coll_{ij}^{\dagger} \Coll_{ij} - \Coll_{ij}^{\dagger} \Coll_{ij} \DensityMatrix(\Time )   }, \nonumber \\
    \Coll_{i1} &= \sqrt{\PhotonLossCoeff \rounds{1+\ThermalOccupation}} \Annihilation_i, \ \ \Coll_{i2} =  \sqrt{\PhotonLossCoeff \ThermalOccupation} \Creation_i, \ \ \Coll_{i3} = \sqrt{\DephasingCoeff} \Creation_i \Annihilation_i,
    \label{full-master}
\end{align}
where collapse operators $\Coll_{i1}$, $\Coll_{i2}$, $\Coll_{i3}$ correspond to deexcitation, excitation and dephasing, respectively, $\PhotonLossCoeff$ and $\DephasingCoeff$ are respective coupling strengths, $\ThermalOccupation = 0$ models pure loss, while $\ThermalOccupation \gg 1$ the heating. 
Such a model captures decoherence processes in mechanical systems---both clamped~\cite{Yang2024} and levitated~\cite{Gonzalez-Ballestero2019}---as well as in optical and microwave cavities~\cite{Liu2026} and trapped ions~\cite{Matsos2025}.
First, we analyze the susceptibility of the checkerboard states against the loss and the heating in the small-decoherence regime, $ \PhotonLossCoeff \Time \ll 1$ and $ \SmallTime \ThermalOccupation \ll 1$~\cite{Fujii2013}. 
Additionally, we assume that noise in the state still allows metrological gain, which is implied by the limits $\PhotonLossCoeff \Time  / \PhaseSpaceDispAmp^2 \ll 1$ and  $\PhotonLossCoeff \Time  \ThermalOccupation / \PhaseSpaceDispAmp^2 \ll 1$ for the loss and heating cases, respectively.
After explicit expansion (see SM~\cite{SuppMat5}) in these respective small parameters, we find for the loss,
\begin{align}
    \CFisher \rounds{\DensityMatrix_\DispLetter}  =  4 \sum_i ( \langle{\NumberOperator_i} \rangle +1) + 4 (1-\frac{\PhotonLossCoeff \Time }{\PhaseSpaceDispAmp^2}) \sum_i \langle{\NumberOperator_i} \rangle +\MMCFisher(\DensityMatrix_\PhaseSpaceDispAmp), 
\end{align}
and for the heating,
\begin{align}
    \CFisher \rounds{\DensityMatrix_\DispLetter}  =  (1-\frac{\PhotonLossCoeff \Time  \ThermalOccupation }{\PhaseSpaceDispAmp^2}) \SMCFisher(\DensityMatrix_\PhaseSpaceDispAmp) +\MMCFisher(\DensityMatrix_\PhaseSpaceDispAmp).
\end{align}
Notably, the cross term is not affected in both cases in this small-decoherence limit, unlike the single-mode term.
The effect of loss can be, in principle, alleviated by error correction~\cite{Noh2020,Zhuang2020} or quantum repeaters~\cite{Xia2019}.
As for the dephasing, one can compute its effect beyond the small-time limit, as the off-diagonal terms in the Fock basis decay exponentially, $\DensityMatrixElement_{nm} (\Time) = \DensityMatrixElement_{nm} \exp[-\DephasingCoeff(n-m)^2 \Time/2]$, yielding
\begin{align}
   \CFisher \rounds{\DensityMatrix_\DispLetter}= \SMCFisher(\DensityMatrix_\PhaseSpaceDispAmp) + e^{-\DephasingCoeff \Time} \MMCFisher(\DensityMatrix_\PhaseSpaceDispAmp).
\end{align}
Expectedly, only intermode coherences are affected.
The last type of noise we consider is phase jitter between the modes during a single experimental shot, which can stem from phase delays between the modes, inconsistent physical orientation of the modes, different couplings of the underlying force source to different modes, and others.
In this case, the FI is
\begin{align}
   \CFisher \rounds{\DensityMatrix_\DispLetter}= \SMCFisher(\DensityMatrix_\PhaseSpaceDispAmp) + e^{-\stdev_{\DispPhase}^2} \MMCFisher(\DensityMatrix_\PhaseSpaceDispAmp),
\end{align}
where $\stdev_{\DispPhase}$ is the strength of the phase jitter (for details see App.~\hyperref[appB]{B}).
Similarly to the dephasing, only the cross term is affected.
The varying robustness of self and cross contributions against different types of noise can therefore guide specific experimental platforms towards either distributed or independent sensing schemes.
Although they have the same noiseless sensitivity, the separable and correlated probe-state strategies from the previous section differ in both noise susceptibility and required resources.
When an experimental setup is dominated by loss or heating, the correlated state performs better; the opposite holds for dephasing and phase jitter.

\textit{Experimental implementation}---The proposed protocol naturally fits platforms with controllable bosonic modes, including microwave cavities in circuit-QED systems and motional modes in circuit-quantum-acoustodynamical (QAD) and trapped-ion systems.
Beyond their maturity for quantum information processing, these platforms are extensively developed for sensing extremely weak signals such as high-frequency gravitational waves~\cite{Fischer2025}, dark matter, fifth forces and other beyond–standard-model phenomena~\cite{Sikivie1983,Teufel2009,Backes2021,Brady2022,Higgins2024,Freiman2025}.
They support coherent displacements, precise multimode coupling, and high-fidelity bosonic readout, making them well-suited for phase-insensitive displacement and force sensing.
While the number of excitations per mode is limited by finite lifetimes, nonlinearities, and measurement resolution, access to multiple modes enables distributed sensing strategies that spread quantum resources across many weakly populated modes, as in our protocol.
High-precision beam-splitter interactions between modes \cite{Gao2018a,vonLupke2024,Qiao2023} enable the preparation of correlated multimode states exploiting collective degrees of freedom and respecting local energy constraints.

The required measurements are experimentally realistic.
Joint parity measurements across multiple bosonic modes have been demonstrated in both circuit-QED~\cite{Riste2013,Huai2024,Hinderling2024} and trapped-ion~\cite{Leibfried1996,Monroe1996,Jia2022,Jeon2025} systems using dispersively coupled ancillae or collective spin–motion mappings, establishing the feasibility of our protocol.
In circuit-QED and -QAD, the mean excitation number per mode is constrained by finite-time state preparation, Kerr nonlinearities inherited from qubits, bosonic enhancement of decay rates, and higher-order corrections to dispersive readout from the Jaynes–Cummings Hamiltonian, while the total number of accessible modes is limited by available volume.
Under these constraints, circuit-QED experiments have achieved $\langle \NumberOperator_i \rangle \sim 1000$ with $M \sim 10$ modes \cite{Milul2023,Huang2025}, whereas circuit-QAD experiments have reached $\langle \NumberOperator_i \rangle \sim 10$ and can support up to $M \sim 10^2$ modes in compact volumes due to the slow speed of sound \cite{Rahman2025}.

\textit{Conclusions and outlook}---To conclude, we have connected phase-insensitive force sensing in continuous-variable systems with distributed quantum sensing, deriving fundamental upper bounds on the achievable amplitude sensitivity when the displacement phase varies randomly between experimental runs.
We identified classes of multimode entangled states and corresponding measurements that saturate these bounds, finding strategies varying in resources needed for the probe state preparation.
Moreover, we analyzed how different decoherence mechanisms affect their performance, revealing distinct scaling laws that can guide experimental implementations.
Our findings are directly relevant for a broad range of continuous-variable platforms---such as trapped ions~\cite{Millican2025}, bulk acoustic~\cite{vonLupke2024} and optomechanical~\cite{Mercade2023,Madiot2023} resonators, circuit-QED devices~\cite{Chakram2021}, and levitated mechanical systems~\cite{Piotrowski2023}---where establishing a shared phase reference is challenging.
Future directions include the simultaneous treatment of phase randomization and stochastic signal amplitudes~\cite{Oh2024,Gardner2025}, exploring whether random states with definite parity offer metrological advantages~\cite{Oszmaniec2016,Kwon2022}, and investigating realistic state-preparation and measurement protocols, potentially optimized by quantum control techniques~\cite{Grochowski2025a,Grochowski2025} and measurement-after-interaction strategies \cite{Guo2025}.
Extending this framework to novel massive mechanical sensors, including weakly nonlinear ones~\cite{Millen2020,Gonzalez-Ballestero2021,Grochowski2025a,Roda-Llordes2024b,Casulleras2024,Rosiek2024}, may further broaden the applicability of phase-insensitive distributed quantum metrology.

\begin{acknowledgments}
\textit{Acknowledgments}---P.T.G. was supported by the project CZ.02.01.01/00/22\_008/0004649 (QUEENTEC) of the EU and the MEYS of the Czech Republic, and project CZ.02.01.01/00/22\_010/0013054 (C-MONS) within Programme JAC MSCA Fellowships at Palacký University Olomouc IV.
M.F. was supported by the Swiss National Science Foundation Ambizione Grant No. 208886, and by The Branco Weiss Fellowship -- Society in Science, administered by the ETH Z\"{u}rich. 
R.F. was supported by the project CZ.02.01.01/00/22\_008/0004649 (QUEENTEC) of the EU and the MEYS of the Czech Republic, project 23-06308S of the Czech Science Foundation, project LUC25006 from the MEYS of the Czech Republic, and COST Action CA23130 (BridgeQG) of the EU.
\end{acknowledgments}

\bibliography{ForceSensing}


\onecolumngrid

\vspace*{5px}
\begin{center}\textbf{ \large End Matter} \end{center}
\vspace*{5px}
\twocolumngrid

\renewcommand{\theequation}{A\arabic{equation}}
\setcounter{equation}{0}

\phantomsection
\label{appA}
\textit{Appendix A: Comparison with a phase-fixed distributed sensing}---Here we check what advantage is brought by usually experimentally challenging fixing the phase between the state and measured phase-space displacement in every experimental shot.
Without loss of generality, we can choose $\DispPhase = \pi/2$ and identify additional contribution to QFI from before phase randomization in Eq.~\eqref{before_mixing},
\begin{align}
 \Delta \QFisher^{\text{pf}} \rounds{\ket{\WaveFunction_\PhaseSpaceDispAmp}} &= 4    \langle(\sum_i \Creation_i )^2\rangle + 4 \langle(\sum_i \Annihilation_i )^2\rangle  = 4 \ModeNumber \langle \CommAnnihilation^{\dagger 2}+ \CommAnnihilation^{ 2}  \rangle \nonumber \\  
 & \leq 8 \ModeNumber \sqrt{\langle \FullNOp\rangle (\langle \FullNOp\rangle+1)},
\end{align}
where the inequality follows from $\langle \CommAnnihilation^{\dagger 2}+ \CommAnnihilation^{ 2}  \rangle/2 = \text{Re} \langle \CommAnnihilation^{\dagger 2} \rangle \leq |\langle \CommAnnihilation^{\dagger 2} \rangle | \leq \sqrt{ \langle \CommCreation \CommAnnihilation\rangle (\langle \CommAnnihilation^\dagger \CommAnnihilation\rangle+1)}$, and $\langle \CommCreation \CommAnnihilation\rangle \leq \langle \FullNOp \rangle = \langle \sum_i \Creation_i \Annihilation_i\rangle$.
Combining with~\eqref{asymbound}, we get the full bound for the force sensing with the fixed phase between the probe and the signal,
\begin{align}
    \QFisher \rounds{\ket{\WaveFunction_\PhaseSpaceDispAmp}} \leq 4 \ModeNumber \left(2 \langle \FullNOp \rangle+1 + 2 \sqrt{\langle \FullNOp\rangle (\langle \FullNOp\rangle+1)} \right),
     \label{asymbound3}
\end{align}
which has been previously reported and is saturated by passively separable states~\cite{Kwon2022}.
Notably, in large-$\langle \FullNOp \rangle$ limit, the scaling $\QFisher \sim \ModeNumber \langle \FullNOp \rangle$ is the same for both phase-randomized and phase-fixed cases, showcasing that phase locking is not a necessary ingredient for the metrological advantage.

\renewcommand{\theequation}{B\arabic{equation}}
\setcounter{equation}{0}

\phantomsection
\label{appB}
\textit{Appendix B: Loss of CFI due to phase jitter}---We model this contribution via averaging over small corrections to the phase-space displacement, $\{\delta \DispPhase_i\}$, which we consider independent in each mode,
\begin{align}
    \DensityMatrix_\DispLetter =&{\int_{0}^{2 \pi}} \dd \delta \DispPhase_1 \dots {\int_{0}^{2 \pi}}\dd \delta \DispPhase_\ModeNumber \probdist_1(\delta \DispPhase_1) \dots  \probdist_\ModeNumber(\delta \DispPhase_\ModeNumber) \times \nonumber \\ & \int_{0}^{2 \pi} \frac{\dd \DispPhase}{2 \pi} 
    \Displacement{\PhaseSpaceDispAmp,\DispPhase, \{\delta \DispPhase_i\}} \DensityMatrix \hat{D}^{\dagger}({\PhaseSpaceDispAmp,\DispPhase, \{\delta \DispPhase_i\}}),
\end{align}
where $\Displacement{\PhaseSpaceDispAmp,\DispPhase, \{\delta \DispPhase_i\}} = \bigotimes_{i=1}^\ModeNumber \hat{D}_i\rounds{\PhaseSpaceDispAmp,\DispPhase+ \delta \DispPhase_i}$.
We assume that each of these corrections follows the normal distribution, $\probdist_i(\delta \DispPhase_i) = \probdist(\delta \DispPhase) \sim \exp(-\delta \DispPhase^2 / 2 \stdev_{\DispPhase}^2 )$.
Via an explicit calculation (for details see SM~\cite{SuppMat5}), the FI reads
\begin{align}
   \CFisher \rounds{\DensityMatrix_\DispLetter}= \SMCFisher(\DensityMatrix_\PhaseSpaceDispAmp) + e^{-\stdev_{\DispPhase}^2} \MMCFisher(\DensityMatrix_\PhaseSpaceDispAmp).
\end{align}


\clearpage

\title{Distributed Phase-Insensitive Displacement Sensing: Supplemental Material}
\maketitle

\makeatletter
\renewcommand{\c@secnumdepth}{0}
\makeatother

\renewcommand{\thesection}{S\Roman{section}}
\renewcommand{\thesubsection}{S\Roman{section}.\arabic{subsection}}
\renewcommand{\thesubsubsection}{S\thesubsection.\arabic{subsubsection}}

\renewcommand{\thefigure}{S\arabic{figure}}
\renewcommand{\theequation}{S\arabic{equation}}

\makeatletter
\renewcommand{\p@subsection}{}
\renewcommand{\p@subsubsection}{}
\renewcommand{\p@figure}{}
\makeatother

\setcounter{section}{0}
\setcounter{equation}{0}
\setcounter{figure}{0}

\onecolumngrid
\section{Saturation via excitation-resolving and total parity measurements}
In this section, we derive the classical Fisher information for both joint-parity and full excitation-resolving measurements by explicit expansion in the Fock basis.
Such a calculation will make the analysis of decoherence possible, which is presented in later sections.
Let us start by explicitly writing~\eqref{mixing} from the main text,
\begin{align}
    \DensityMatrix_\DispLetter =\int_{0}^{2 \pi} \frac{\dd \DispPhase}{2 \pi} \sum_{k_1,l_1,k_2,l_2,\dots,k_\ModeNumber,l_\ModeNumber} \DensityMatrixElement_{k_1,l_1,k_2,l_2,\dots,k_\ModeNumber,l_\ModeNumber} \bigotimes_{i=1}^\ModeNumber \hat{D}_i\rounds{\PhaseSpaceDispAmp,\DispPhase} \ket{k_i}_i\bra{l_i} \hat{D}_i^{\dagger}\rounds{\PhaseSpaceDispAmp,\DispPhase}.
    \label{full_state_mix}
\end{align}
Let consider small $\PhaseSpaceDispAmp$ expansion,
\begin{align}
    \hat{D}_i\rounds{\PhaseSpaceDispAmp,\DispPhase} \ket{k_i}_i =& \ket{k_i}_i + \PhaseSpaceDispAmp e^{\ImagUnit \DispPhase} \sqrt{k_i+1} \ket{k_i+1}_i - \PhaseSpaceDispAmp e^{-\ImagUnit \DispPhase} \sqrt{k_i} \ket{k_i-1}_i + \nonumber \\
    & \frac{1}{2} \PhaseSpaceDispAmp^2 e^{\ImagUnit 2 \DispPhase} \sqrt{k_i+1}\sqrt{k_i+2} \ket{k_i+2}_i + \frac{1}{2} \PhaseSpaceDispAmp^2 e^{-\ImagUnit 2 \DispPhase} \sqrt{k_i}\sqrt{k_i-1} \ket{k_i-2}_i - \nonumber \\ 
    & \frac{1}{2} \PhaseSpaceDispAmp^2 \rounds{2k_i+1} \ket{k_i}_i 
    +\mathcal{O}\rounds{\PhaseSpaceDispAmp^3}.
\end{align}
Then, we can write
\begin{align}
    \hat{D}_i\rounds{\PhaseSpaceDispAmp,\DispPhase} \ket{k_i}_i\bra{l_i} \hat{D}_i^{\dagger}\rounds{\PhaseSpaceDispAmp,\DispPhase} =& \ket{k_i}_i\bra{l_i}  + \PhaseSpaceDispAmp e^{\ImagUnit \DispPhase} \sqrt{k_i+1} \ket{k_i+1}_i\bra{l_i} - \PhaseSpaceDispAmp e^{-\ImagUnit \DispPhase} \sqrt{k_i} \ket{k_i-1}_i\bra{l_i} + \nonumber \\ 
     &\PhaseSpaceDispAmp e^{-\ImagUnit \DispPhase} \sqrt{l_i+1} \ket{k_i}_i\bra{l_i+1} - \PhaseSpaceDispAmp e^{\ImagUnit \DispPhase} \sqrt{l_i} \ket{k_i}_i\bra{l_i-1} +\nonumber \\ &\PhaseSpaceDispAmp^2 \sqrt{k_i+1}\sqrt{l_i+1} \ket{k_i+1}_i\bra{l_i+1} + \PhaseSpaceDispAmp^2 \sqrt{k_i-1}\sqrt{l_i-1} \ket{k_i-1}_i\bra{l_i-1}- \nonumber \\
     & \PhaseSpaceDispAmp^2 e^{-\ImagUnit 2\DispPhase} \sqrt{k_i}\sqrt{l_i+1} \ket{k_i-1}_i\bra{l_i+1} - \PhaseSpaceDispAmp^2 e^{\ImagUnit 2\DispPhase}  \sqrt{k_i+1}\sqrt{l_i} \ket{k_i+1}_i\bra{l_i-1}  +\nonumber \\
     &\frac{1}{2} \PhaseSpaceDispAmp^2 e^{-\ImagUnit 2\DispPhase} \sqrt{l_i+1}\sqrt{l_i+2}  \ket{k_i}_i\bra{l_i+2} +  \frac{1}{2} \PhaseSpaceDispAmp^2 e^{\ImagUnit 2\DispPhase} \sqrt{l_i}\sqrt{l_i-1}  \ket{k_i}_i\bra{l_i-2} +\nonumber \\
     &\frac{1}{2} \PhaseSpaceDispAmp^2 e^{\ImagUnit 2\DispPhase} \sqrt{k_i+1}\sqrt{k_i+2}  \ket{k_i+2}_i\bra{l_i} +  \frac{1}{2} \PhaseSpaceDispAmp^2 e^{-\ImagUnit 2\DispPhase} \sqrt{k_i}\sqrt{k_i-1}  \ket{k_i-2}_i\bra{l_i}  -\nonumber \\
     & \frac{1}{2}\PhaseSpaceDispAmp^2 \rounds{2 k_i + 2 l_i + 2} \ket{k_i}_i\bra{l_i} +\mathcal{O}\rounds{\PhaseSpaceDispAmp^3},
\end{align}
and
\begin{align}
    &\int_{0}^{2 \pi} \frac{\dd \DispPhase}{2 \pi} \bigotimes_{i=1}^\ModeNumber \hat{D}_i\rounds{\PhaseSpaceDispAmp,\DispPhase} \ket{k_i}_i\bra{l_i} \hat{D}_i^{\dagger}\rounds{\PhaseSpaceDispAmp,\DispPhase} = \bigotimes_{i=1}^\ModeNumber \ket{k_i}_i\bra{l_i} \squares{1-\frac{\PhaseSpaceDispAmp^2}{2} \sum_{j=1}^{\ModeNumber} \rounds{2 k_j + 2 l_j + 2} }+ \nonumber \\
    & \PhaseSpaceDispAmp^2 \sum_{j,j'} \ket{j,j'}\bra{j,j'} \Big[
    \sqrt{k_{j'}+1} \sqrt{l_{j'}+1} \ket{k_{j}}_{j}\bra{l_{j}} \ket{k_{j'}+1}_{j'}\bra{l_{j'}+1} + 
     \sqrt{k_{j'}} \sqrt{l_{j'}} \ket{k_{j}}_{j}\bra{l_{j}} \ket{k_{j'}-1}_{j'}\bra{l_{j'}-1} +
    \nonumber \\
    &\sqrt{k_{j}+1} \sqrt{l_{j}+1} \ket{k_{j}+1}_{j}\bra{l_{j}+1} \ket{k_{j'}}_{j'}\bra{l_{j'}} + 
     \sqrt{k_{j}} \sqrt{l_{j}} \ket{k_{j}-1}_{j}\bra{l_{j}-1} \ket{k_{j'}}_{j'}\bra{l_{j'}} -    
     \nonumber \\
    &\sqrt{k_{j}+1} \sqrt{k_{j'}} \ket{k_{j}+1}_{j}\bra{l_{j}} \ket{k_{j'}-1}_{j'}\bra{l_{j'}} + 
     \sqrt{k_{j}+1} \sqrt{l_{j'}+1} \ket{k_{j}+1}_{j}\bra{l_{j}} \ket{k_{j'}}_{j'}\bra{l_{j'}+1} -    
     \nonumber \\
    &\sqrt{k_{j}} \sqrt{k_{j'}+1} \ket{k_{j}-1}_{j}\bra{l_{j}} \ket{k_{j'}+1}_{j'}\bra{l_{j'}} + 
     \sqrt{k_{j}} \sqrt{l_{j'}} \ket{k_{j}-1}_{j}\bra{l_{j}} \ket{k_{j'}}_{j'}\bra{l_{j'}-1} +    
     \nonumber \\  
    &\sqrt{l_{j}+1} \sqrt{k_{j'}+1} \ket{k_{j}}_{j}\bra{l_{j}+1} \ket{k_{j'}+1}_{j'}\bra{l_{j'}} - 
     \sqrt{l_{j}+1} \sqrt{l_{j'}} \ket{k_{j}}_{j}\bra{l_{j}+1} \ket{k_{j'}}_{j'}\bra{l_{j'}-1} +    
     \nonumber \\
    &\sqrt{l_{j}} \sqrt{k_{j'}} \ket{k_{j}}_{j}\bra{l_{j}-1} \ket{k_{j'}-1}_{j'}\bra{l_{j'}} - 
     \sqrt{l_{j}} \sqrt{l_{j'}+1} \ket{k_{j}}_{j}\bra{l_{j}-1} \ket{k_{j'}}_{j'}\bra{l_{j'}+1}\Big] +\mathcal{O}\rounds{\PhaseSpaceDispAmp^3},
    \label{single_braket}
\end{align}
where
\begin{align}
 \ket{j,j'}\bra{j,j'} = \ket{k_1}_1\bra{l_1}\dots \ket{k_{j-1}}_{j-1}\bra{l_{j-1}}\ket{k_{j+1}}_{j+1}\bra{l_{j+1}} \dots \ket{k_{j'-1}}_{j'-1}\bra{l_{j'-1}}\ket{k_{j'+1}}_{j'+1}\bra{l_{j'+1}}\dots \ket{k_\ModeNumber}_\ModeNumber\bra{l_\ModeNumber}.
\end{align}
First, we start with the excitation-resolving measurement.
We consider the measurement of only classical correlations between the modes, which are encoded in the multimode probability distribution
\begin{align}
 \Prob\rounds{n_1,\dots,n_\ModeNumber} = \bigotimes_{i=1}^\ModeNumber \bra{n_i} \DensityMatrix_\DispLetter \bigotimes_{i=1}^\ModeNumber \ket{n_i}.\label{multimode}
\end{align}
After using Eqs.~\eqref{full_state_mix},~\eqref{single_braket}, and~\eqref{multimode}, we can write
\begin{align}
 \Prob_{\PhaseSpaceDispAmp}\rounds{n_1,\dots,n_\ModeNumber} = & \DensityMatrixElement_{n_1,n_1,n_2,n_2,\dots,n_\ModeNumber,n_\ModeNumber} \squares{1-\PhaseSpaceDispAmp^2 \sum_{j=1}^{\ModeNumber} \rounds{2 n_j + 1} } + \nonumber \\
 \PhaseSpaceDispAmp^2 \sum_{j,j'} \Big[ & n_{j'} \DensityMatrixElement_{n_{j},n_{j},n_{j'}-1,n_{j'}-1}^{n_1,\dots,n_\ModeNumber;{j,j'}}  +
 (n_{j'}+1) \DensityMatrixElement_{n_{j},n_{j},n_{j'}+1,n_{j'}+1}^{n_1,\dots,n_\ModeNumber;{j,j'}}
 + \nonumber \\
& n_{j} \DensityMatrixElement_{n_{j}-1,n_{j}-1,n_{j'},n_{j'}}^{n_1,\dots,n_\ModeNumber;{j,j'}}  +
 (n_{j}+1) \DensityMatrixElement_{n_{j}+1,n_{j}+1,n_{j'},n_{j'}}^{n_1,\dots,n_\ModeNumber;{j,j'}} -\nonumber \\
 & \sqrt{n_j} \sqrt{n_{j'}+1} \DensityMatrixElement_{n_{j}-1,n_{j},n_{j'}+1,n_{j'}}^{n_1,\dots,n_\ModeNumber;{j,j'}} + \sqrt{n_j} \sqrt{n_{j'}} \DensityMatrixElement_{n_{j}-1,n_{j},n_{j'},n_{j'}-1}^{n_1,\dots,n_\ModeNumber;{j,j'}} - \nonumber \\
 &\sqrt{n_j+1} \sqrt{n_{j'}} \DensityMatrixElement_{n_{j}+1,n_{j},n_{j'}-1,n_{j'}}^{n_1,\dots,n_\ModeNumber;{j,j'}} + \sqrt{n_j+1} \sqrt{n_{j'}+1} \DensityMatrixElement_{n_{j}+1,n_{j},n_{j'},n_{j'}+1}^{n_1,\dots,n_\ModeNumber;{j,j'}} + \nonumber \\
 & \sqrt{n_j} \sqrt{n_{j'}} \DensityMatrixElement_{n_{j},n_{j}-1,n_{j'}-1,n_{j'}}^{n_1,\dots,n_\ModeNumber;{j,j'}} - \sqrt{n_j} \sqrt{n_{j'}+1} \DensityMatrixElement_{n_{j},n_{j}-1,n_{j'},n_{j'}+1}^{n_1,\dots,n_\ModeNumber;{j,j'}} + \nonumber \\
 &\sqrt{n_j+1} \sqrt{n_{j'}+1} \DensityMatrixElement_{n_{j},n_{j}+1,n_{j'}+1,n_{j'}}^{n_1,\dots,n_\ModeNumber;{j,j'}} - \sqrt{n_j+1} \sqrt{n_{j'}} \DensityMatrixElement_{n_{j},n_{j}+1,n_{j'},n_{j'}-1}^{n_1,\dots,n_\ModeNumber;{j,j'}} 
 \Big] + \mathcal{O}\rounds{\PhaseSpaceDispAmp^3},
 \label{pnin}
\end{align}
where
\begin{align}
\DensityMatrixElement_{k,l,k',l'}^{n_1,\dots,n_\ModeNumber;{j,j'}}  = \DensityMatrixElement_{n_1,n_1,\dots,n_{j-1},n_{j-1},k,l,n_{j+1},n_{j+1},\dots,n_{j'-1},n_{j'-1},k',l',n_{j'+1},n_{j'+1},\dots,n_\ModeNumber,n_\ModeNumber}.
\end{align}
The subscript after the semicolon denotes which entries from $(n_1,\dots,n_\ModeNumber)$ are altered.
Let us now list all the elements $ \Prob_\PhaseSpaceDispAmp\rounds{n_1,\dots,n_\ModeNumber} $ that contain specific density matrix entries $\DensityMatrixElement_{k,l,k',l'}^{m_1,\dots,m_\ModeNumber;{j,j'}} $ for a fixed $(m_1,\dots,m_\ModeNumber)$.
First, let us denote 
\begin{align}
\DensityMatrixElement^{m_1,\dots,m_\ModeNumber}_{j,j'}  &= \DensityMatrixElement_{m_j,m_j,m_{j'},m_{j'}}^{m_1,\dots,m_\ModeNumber;{j,j'}}, \nonumber \\
\DensityMatrixElement^{m_1,\dots,m_\ModeNumber}_{\text{R};j,j'}  &= \DensityMatrixElement_{m_j,m_j+1,m_{j'},m_{j'}-1}^{m_1,\dots,m_\ModeNumber;{j,j'}}, \nonumber \\
\DensityMatrixElement^{m_1,\dots,m_\ModeNumber}_{\text{L};j,j'}  &= \DensityMatrixElement_{m_j,m_j-1,m_{j'},m_{j'}+1}^{m_1,\dots,m_\ModeNumber;{j,j'}}.
\label{entries}
\end{align}
Then, for each pair $j \neq j'$, we have 7 affected  $ \Prob_\PhaseSpaceDispAmp\rounds{n_1,\dots,n_\ModeNumber} $ in the order $\mathcal{O}(\PhaseSpaceDispAmp^2)$;
\begin{align}
\Prob_\PhaseSpaceDispAmp\rounds{m_1,\dots,m_j+1,\dots,m_{j'}\dots,m_\ModeNumber}  &= \Prob_0\rounds{m_1,\dots,m_j+1,\dots,m_{j'}\dots,m_\ModeNumber} \nonumber \\
& \quad \quad \quad \quad \quad \quad \quad+\PhaseSpaceDispAmp^2 (m_j+1) \DensityMatrixElement^{m_1,\dots,m_\ModeNumber}_{j,j'}  + \PhaseSpaceDispAmp^2 \sqrt{m_j+1} \sqrt{m_{j'}}\DensityMatrixElement^{m_1,\dots,m_\ModeNumber}_{\text{R};j,j'}, \nonumber \\
\Prob_\PhaseSpaceDispAmp\rounds{m_1,\dots,m_j-1,\dots,m_{j'}\dots,m_\ModeNumber}  &= \Prob_0\rounds{m_1,\dots,m_j-1,\dots,m_{j'}\dots,m_\ModeNumber}\nonumber \\
& \quad \quad \quad \quad \quad \quad \quad + \PhaseSpaceDispAmp^2 m_j \DensityMatrixElement^{m_1,\dots,m_\ModeNumber}_{j,j'}  + \PhaseSpaceDispAmp^2 \sqrt{m_j} \sqrt{m_{j'}+1}\DensityMatrixElement^{m_1,\dots,m_\ModeNumber}_{\text{L};j,j'},
\nonumber \\
\Prob_\PhaseSpaceDispAmp\rounds{m_1,\dots,m_j,\dots,m_{j'}+1\dots,m_\ModeNumber}  &= \Prob_0\rounds{m_1,\dots,m_j,\dots,m_{j'}+1\dots,m_\ModeNumber}\nonumber \\
& \quad \quad \quad \quad \quad \quad \quad + \PhaseSpaceDispAmp^2 (m_{j'}+1) \DensityMatrixElement^{m_1,\dots,m_\ModeNumber}_{j,j'}  + \PhaseSpaceDispAmp^2 \sqrt{m_j} \sqrt{m_{j'}+1}\DensityMatrixElement^{m_1,\dots,m_\ModeNumber}_{\text{L};j,j'},
\nonumber \\
\Prob_\PhaseSpaceDispAmp\rounds{m_1,\dots,m_j,\dots,m_{j'}-1\dots,m_\ModeNumber}  &= \Prob_0\rounds{m_1,\dots,m_j,\dots,m_{j'}-1\dots,m_\ModeNumber} \nonumber \\
& \quad \quad \quad \quad \quad \quad \quad+ \PhaseSpaceDispAmp^2 m_{j'} \DensityMatrixElement^{m_1,\dots,m_\ModeNumber}_{j,j'}  + \PhaseSpaceDispAmp^2 \sqrt{m_j+1} \sqrt{m_{j'}}\DensityMatrixElement^{m_1,\dots,m_\ModeNumber}_{\text{R};j,j'},
\nonumber \\
\Prob_\PhaseSpaceDispAmp\rounds{m_1,\dots,m_j,\dots,m_{j'}\dots,m_\ModeNumber}  &= \Prob_0\rounds{m_1,\dots,m_j,\dots,m_{j'}\dots,m_\ModeNumber} \squares{1-\PhaseSpaceDispAmp^2 \sum_{i=1}^{\ModeNumber} \rounds{2 n_i + 1} } \nonumber \\
& \quad \quad \quad \quad \quad \quad \quad - \PhaseSpaceDispAmp^2 \sqrt{m_j+1} \sqrt{m_{j'}} \DensityMatrixElement^{m_1,\dots,m_\ModeNumber}_{\text{R};j,j'}  - \PhaseSpaceDispAmp^2 \sqrt{m_j} \sqrt{m_{j'}+1}\DensityMatrixElement^{m_1,\dots,m_\ModeNumber}_{\text{R};j,j'},
\nonumber \\
\Prob_\PhaseSpaceDispAmp\rounds{m_1,\dots,m_j+1,\dots,m_{j'}-1\dots,m_\ModeNumber}  &= \Prob_0\rounds{m_1,\dots,m_j+1,\dots,m_{j'}-1\dots,m_\ModeNumber} -\PhaseSpaceDispAmp^2 \sqrt{m_j+1} \sqrt{m_{j'}}\DensityMatrixElement^{m_1,\dots,m_\ModeNumber}_{\text{R};j,j'},
\nonumber \\
\Prob_\PhaseSpaceDispAmp\rounds{m_1,\dots,m_j-1,\dots,m_{j'}+1\dots,m_\ModeNumber}  &= \Prob_0\rounds{m_1,\dots,m_j-1,\dots,m_{j'}+1\dots,m_\ModeNumber} -\PhaseSpaceDispAmp^2 \sqrt{m_j} \sqrt{m_{j'}+1}\DensityMatrixElement^{m_1,\dots,m_\ModeNumber}_{\text{L};j,j'}.\label{listing}
\end{align}

Let us now switch attention to the classical Fisher information for such a measurement:
\begin{align}
    \Fisher_\PhaseSpaceDispAmp = \sum_{n_1,\dots,n_\ModeNumber} \frac{\squares{\partial_\PhaseSpaceDispAmp \Prob_\PhaseSpaceDispAmp\rounds{n_1,\dots,n_\ModeNumber}}^2}{\Prob_\PhaseSpaceDispAmp\rounds{n_1,\dots,n_\ModeNumber}} = \sum_{n_1,\dots,n_\ModeNumber} \SmallFisher_\PhaseSpaceDispAmp\rounds{n_1,\dots,n_\ModeNumber}  .
\end{align}
First, note that if $\Prob_0\rounds{n_1,\dots,n_\ModeNumber} \neq 0$, then
\begin{align}
    \Prob_\PhaseSpaceDispAmp\rounds{n_1,\dots,n_\ModeNumber} = \Prob_0\rounds{n_1,\dots,n_\ModeNumber} + \beta \PhaseSpaceDispAmp^r +  \mathcal{O}(\PhaseSpaceDispAmp^{r+1} ),
\end{align}
and
\begin{align}
    \SmallFisher_\PhaseSpaceDispAmp\rounds{n_1,\dots,n_\ModeNumber} = \frac{r^2 \PhaseSpaceDispAmp^{2(r-1)} \beta^2}{\Prob_0\rounds{n_1,\dots,n_\ModeNumber} } +   \mathcal{O}(\PhaseSpaceDispAmp^{2(r-1)+1} ),
\end{align}
which vanishes in the limit $\PhaseSpaceDispAmp \rightarrow 0$.
However, when $\Prob_0\rounds{n_1,\dots,n_\ModeNumber} = 0$, then 
\begin{align}
    \Prob_\PhaseSpaceDispAmp\rounds{n_1,\dots,n_\ModeNumber} =  \beta  \PhaseSpaceDispAmp^r +  \mathcal{O}(\PhaseSpaceDispAmp^{r+1} ),
\end{align}
and
\begin{align}
    \SmallFisher_\PhaseSpaceDispAmp\rounds{n_1,\dots,n_\ModeNumber} = r^2 \PhaseSpaceDispAmp^{r-2} \beta +   \mathcal{O}(\PhaseSpaceDispAmp^{r+1} ).\label{fish_cont}
\end{align}
Then, only for $r=2$ we have a nonzero contribution to the Fisher information in the limit $\PhaseSpaceDispAmp \rightarrow 0$.
Hence, to get the nonzero Fisher information, we need to look at the states for which contributions~\eqref{listing} to specific $\Prob_\PhaseSpaceDispAmp\rounds{n_1,\dots,n_\ModeNumber}$ when $\Prob_0\rounds{n_1,\dots,n_\ModeNumber}$ are of the order $\mathcal{O}(\PhaseSpaceDispAmp^2)$.

The example of such states are the states with the fixed total parity:
\begin{align}
    \Prob_0\rounds{n_1,\dots,n_\ModeNumber} \neq 0 \ \ \text{only if} \ \  \sum_i n_i = 2s+1; \ s \in \mathbb{Z}
\end{align}
for odd-parity states or 
\begin{align}
    \Prob_0\rounds{n_1,\dots,n_\ModeNumber} \neq 0 \ \ \text{only if} \ \  \sum_i n_i = 2s; \ s \in \mathbb{Z}
    \label{even-parity}
\end{align}
for even-parity states.
For brevity, we will call them \textit{checkerboard} states as their occupied Fock states represented on the grid make a high-dimensional checkerboard.

For such states, only the first four terms in~\eqref{listing} contribute.
Specifically, all the $\Prob_0\rounds{n_1,\dots,n_\ModeNumber} = 0$, namely with other parity than the initial one, are to be measured.
Moreover, given~\eqref{fish_cont}, the quadratic corrections to $\Prob_\PhaseSpaceDispAmp\rounds{n_1,\dots,n_\ModeNumber}$ add linearly to the Fisher information, i.e.,
\begin{align}
    \Prob_\PhaseSpaceDispAmp\rounds{n_1,\dots,n_\ModeNumber} = \beta \PhaseSpaceDispAmp^2 \ \ \rightarrow \ \ \SmallFisher_\PhaseSpaceDispAmp\rounds{n_1,\dots,n_\ModeNumber} = 4 \beta, 
\end{align}
where we put only the leading terms.
Then, let us sum up the contributions coming from~\eqref{listing}
\begin{align}
\SmallFisher_\PhaseSpaceDispAmp\rounds{m_1,\dots,m_j+1,\dots,m_{j'}\dots,m_\ModeNumber}  &=  4(m_j+1) \DensityMatrixElement^{m_1,\dots,m_\ModeNumber}_{j,j'}  + 4\sqrt{m_j+1} \sqrt{m_{j'}}\DensityMatrixElement^{m_1,\dots,m_\ModeNumber}_{\text{R};j,j'}, \nonumber \\
\SmallFisher_\PhaseSpaceDispAmp\rounds{m_1,\dots,m_j-1,\dots,m_{j'}\dots,m_\ModeNumber}  &= 4 m_j \DensityMatrixElement^{m_1,\dots,m_\ModeNumber}_{j,j'}  + 4 \sqrt{m_j} \sqrt{m_{j'}+1}\DensityMatrixElement^{m_1,\dots,m_\ModeNumber}_{\text{L};j,j'},
\nonumber \\
\SmallFisher_\PhaseSpaceDispAmp\rounds{m_1,\dots,m_j,\dots,m_{j'}+1\dots,m_\ModeNumber}  &= 4 (m_{j'}+1) \DensityMatrixElement^{m_1,\dots,m_\ModeNumber}_{j,j'}  + 4 \sqrt{m_j} \sqrt{m_{j'}+1}\DensityMatrixElement^{m_1,\dots,m_\ModeNumber}_{\text{L};j,j'},
\nonumber \\
\SmallFisher_\PhaseSpaceDispAmp\rounds{m_1,\dots,m_j,\dots,m_{j'}-1\dots,m_\ModeNumber}  &= 4 m_{j'} \DensityMatrixElement^{m_1,\dots,m_\ModeNumber}_{j,j'}  + 4 \sqrt{m_j+1} \sqrt{m_{j'}}\DensityMatrixElement^{m_1,\dots,m_\ModeNumber}_{\text{R};j,j'},\label{listing2}
\end{align}
where only the leading term is present.
Then, the total Fisher information reads
\begin{align}
   \Fisher_\PhaseSpaceDispAmp =& \frac{1}{2} \sum_{m_1,\dots,m_\ModeNumber}  \sum_{j,j'} \SmallFisher_\PhaseSpaceDispAmp\rounds{m_1,\dots,m_j+1,\dots,m_{j'}\dots,m_\ModeNumber} + \SmallFisher_\PhaseSpaceDispAmp\rounds{m_1,\dots,m_j-1,\dots,m_{j'}\dots,m_\ModeNumber} + \nonumber \\
   & \quad \quad \quad \quad \SmallFisher_\PhaseSpaceDispAmp\rounds{m_1,\dots,m_j,\dots,m_{j'}+1\dots,m_\ModeNumber} + \SmallFisher_\PhaseSpaceDispAmp\rounds{m_1,\dots,m_j,\dots,m_{j'}-1\dots,m_\ModeNumber} \nonumber \\
    =& 4 \sum_{m_1,\dots,m_\ModeNumber}  \sum_{j} \DensityMatrixElement_{m_1,m_1,\dots,m_\ModeNumber,m_\ModeNumber} (2m_i+1) + 4 \sum_{m_1,\dots,m_\ModeNumber}  \sum_{j\neq j'} \rounds{\sqrt{m_j+1} \sqrt{m_{j'}}\DensityMatrixElement^{m_1,\dots,m_\ModeNumber}_{\text{R};j,j'} + \sqrt{m_j} \sqrt{m_{j'}+1}\DensityMatrixElement^{m_1,\dots,m_\ModeNumber}_{\text{L};j,j'}} , 
\end{align}
where we took care of the double counting by the first prefactor.

Let us consider a pure state 
\begin{align}
\ket{\WaveFunction} = \sum_{n_1,\dots,n_\ModeNumber} \WFCoeff_{n_1,\dots,n_\ModeNumber} \ket{n_1}_1 \dots \ket{n_\ModeNumber}_\ModeNumber.
\end{align}
Then, 
\begin{align}
\DensityMatrixElement^{n_1,\dots,n_\ModeNumber}_{j,j'}  &= |\WFCoeff_{n_1,\dots,n_\ModeNumber}|^2, \nonumber \\
\DensityMatrixElement^{n_1,\dots,n_\ModeNumber}_{\text{R};j,j'}  &= \WFCoeff_{n_1,\dots,n_\ModeNumber} \WFCoeff_{n_1,\dots,n_j+1,\dots,n_{j'}-1,\dots,n_\ModeNumber}^*, \nonumber \\
\DensityMatrixElement^{n_1,\dots,n_\ModeNumber}_{\text{L};j,j'}  &= \WFCoeff_{n_1,\dots,n_\ModeNumber} \WFCoeff_{n_1,\dots,n_j-1,\dots,n_{j'}+1,\dots,n_\ModeNumber}^*.
\end{align}
Straightforwardly for $j \neq j'$,
\begin{align}
    \angles{ \Annihilation_{j} \Creation_{j'} } = \sum_{n_1,\dots,n_\ModeNumber} \WFCoeff_{n_1,\dots,n_\ModeNumber} \WFCoeff_{n_1,\dots,n_j+1,\dots,n_{j'}-1,\dots,n_\ModeNumber}^* \sqrt{n_j} \sqrt{n_{j'}+1} = \sum_{n_1,\dots,n_\ModeNumber} \DensityMatrixElement^{n_1,\dots,n_\ModeNumber}_{\text{L};j,j'}\sqrt{n_j} \sqrt{n_{j'}+1}, \nonumber \\
    \angles{ \Annihilation_{j'} \Creation_{j} } = \sum_{n_1,\dots,n_\ModeNumber} \WFCoeff_{n_1,\dots,n_\ModeNumber} \WFCoeff_{n_1,\dots,n_j-1,\dots,n_{j'}+1,\dots,n_\ModeNumber}^* \sqrt{n_{j'}} \sqrt{n_{j}+1} = \sum_{n_1,\dots,n_\ModeNumber} \DensityMatrixElement^{n_1,\dots,n_\ModeNumber}_{\text{R};j,j'}\sqrt{n_{j'}} \sqrt{n_{j}+1},
\end{align}
so we can write based on~\eqref{full_bound},
\begin{align}
    \QFisher \rounds{\DensityMatrix_\DispLetter} = \CFisher^{\text{exc}} \rounds{\DensityMatrix_\DispLetter},
\end{align}
where we denoted Fisher information for the excitation-resolving measurement as $\CFisher^{\text{exc}}$ and the equality holds for pure checkerboard states in the limit $\PhaseSpaceDispAmp \rightarrow 0$.

Let us take a look at another type of measurement, namely the total parity measurement:
\begin{equation}
    \ParityOp = (-1)^{\sum_i \Creation_i \Annihilation_i}.
\end{equation}
The expectation value for a given state for this operator equals
\begin{equation}
    \angles{\ParityOp} = - \sum_{\substack{n_1,\dots,n_\ModeNumber \\ \sum_i n_i = 2s+1; \ s \in \mathbb{Z}}} \DensityMatrixElement_{n_1,n_1,\dots,n_\ModeNumber,n_\ModeNumber} + \sum_{\substack{n_1,\dots,n_\ModeNumber \\ \sum_i n_i = 2s; \ s \in \mathbb{Z}}} \DensityMatrixElement_{n_1,n_1,\dots,n_\ModeNumber,n_\ModeNumber}.
\end{equation}
Let us consider even-parity state~\eqref{even-parity}.
Then, $\angles{\ParityOp} = 1$.
We can extract the effect on this observable from~\eqref{listing}, realizing that the first four terms induce transitions only to opposite-parity states.
The total change then reads
\begin{align}
  \angles{\ParityOp}_\PhaseSpaceDispAmp =1 - &2\PhaseSpaceDispAmp^2 \sum_{m_1,\dots,m_\ModeNumber}  \sum_{j} \DensityMatrixElement_{m_1,m_1,\dots,m_\ModeNumber,m_\ModeNumber} (2m_i+1)\nonumber \\  - &2\PhaseSpaceDispAmp^2 \sum_{m_1,\dots,m_\ModeNumber}  \sum_{j\neq j'} \rounds{\sqrt{m_j+1} \sqrt{m_{j'}}\DensityMatrixElement^{m_1,\dots,m_\ModeNumber}_{\text{R};j,j'} + \sqrt{m_j} \sqrt{m_{j'}+1}\DensityMatrixElement^{m_1,\dots,m_\ModeNumber}_{\text{L};j,j'}}.   
\end{align}
Then, the probability of getting parity 1 reads
\begin{align}
    \Prob_1 = \frac{1}{2} & \rounds{1+ \angles{\ParityOp}_\PhaseSpaceDispAmp } = \nonumber \\  1 - &\PhaseSpaceDispAmp^2 \sum_{m_1,\dots,m_\ModeNumber}  \sum_{j} \DensityMatrixElement_{m_1,m_1,\dots,m_\ModeNumber,m_\ModeNumber} (2m_i+1)  \nonumber \\ - &\PhaseSpaceDispAmp^2 \sum_{m_1,\dots,m_\ModeNumber}  \sum_{j\neq j'} \rounds{\sqrt{m_j+1} \sqrt{m_{j'}}\DensityMatrixElement^{m_1,\dots,m_\ModeNumber}_{\text{R};j,j'} + \sqrt{m_j} \sqrt{m_{j'}+1}\DensityMatrixElement^{m_1,\dots,m_\ModeNumber}_{\text{L};j,j'}},
\end{align}
and the Fisher information
\begin{align}
    \CFisher^{\Parity} \rounds{\DensityMatrix_\DispLetter} = \frac{\rounds{\partial_\PhaseSpaceDispAmp \Prob_1}^2}{\Prob_1 (1-\Prob_1)}  =&  4 \sum_{m_1,\dots,m_\ModeNumber}  \sum_{j} \DensityMatrixElement_{m_1,m_1,\dots,m_\ModeNumber,m_\ModeNumber} (2m_i+1)  +\nonumber \\ &4 \sum_{m_1,\dots,m_\ModeNumber}  \sum_{j\neq j'} \rounds{\sqrt{m_j+1} \sqrt{m_{j'}}\DensityMatrixElement^{m_1,\dots,m_\ModeNumber}_{\text{R};j,j'} + \sqrt{m_j} \sqrt{m_{j'}+1}\DensityMatrixElement^{m_1,\dots,m_\ModeNumber}_{\text{L};j,j'}} = \CFisher^{\text{exc}} \rounds{\DensityMatrix_\DispLetter},
\end{align}
where we didn't explicitly write the order $\mathcal{O}(\PhaseSpaceDispAmp^3)$.

\section{Decoherence}
\subsection{Phase jitter}
Let us now consider how robust the scheme is against the phase jitter, namely, let us consider small stochastic corrections to phases varying between the modes,
\begin{align}
    \DensityMatrix_\DispLetter ={\int_{0}^{2 \pi}}\delta \DispPhase_1 \dots \delta \DispPhase_\ModeNumber \probdist_1(\delta \DispPhase_1) \dots  \probdist_\ModeNumber(\delta \DispPhase_\ModeNumber)  \int_{0}^{2 \pi} \frac{\dd \DispPhase}{2 \pi} &\sum_{k_1,l_1,k_2,l_2,\dots,k_\ModeNumber,l_\ModeNumber} \DensityMatrixElement_{k_1,l_1,k_2,l_2,\dots,k_\ModeNumber,l_\ModeNumber} \times\nonumber \\ &\bigotimes_{i=1}^\ModeNumber \hat{D}_i\rounds{\PhaseSpaceDispAmp,\DispPhase + \delta \DispPhase_i} \ket{k_i}_i\bra{l_i} \hat{D}_i^{\dagger}\rounds{\PhaseSpaceDispAmp,\DispPhase + \delta \DispPhase_i}.
\end{align}
We will assume that each of these corrections follows the normal distribution,
\begin{align}
    \probdist_i(\delta \DispPhase_i) = \probdist(\delta \DispPhase) = \frac{1}{\sqrt{2 \pi \stdev_{\DispPhase}^2}} e^{-\frac{\delta \DispPhase^2}{2 \stdev_{\DispPhase}^2 }}.
\end{align}
In Eq.~\eqref{single_braket}, only the off-diagonal terms, i.e., $q_1 \neq q_2$ or $q_3 \neq q_4$, are affected,
\begin{align}
\ket{k_{j}+q_1}_{j}\bra{l_{j}+q_2}\ket{k_{j'}+q_3}_{j'}\bra{l_{j'}+q_4} \rightarrow e^{\pm \ImagUnit  \delta \DispPhase_j } e^{\pm \ImagUnit  \delta \DispPhase_{j'} } \ket{k_{j}+q_1}_{j}\bra{l_{j}+q_2}\ket{k_{j'}+q_3}_{j'}\bra{l_{j'}+q_4} ,
\end{align}
where each of $\pm$ is independent and depends on specific $q_1,q_2,q_3,q_4$.
After the integration over $\delta \DispPhase_j$, the new Eq.~\eqref{pnin} read
\begin{align}
 &\Prob_{\PhaseSpaceDispAmp}\rounds{n_1,\dots,n_\ModeNumber} = \DensityMatrixElement_{n_1,n_1,n_2,n_2,\dots,n_\ModeNumber,n_\ModeNumber} \squares{1-\PhaseSpaceDispAmp^2 \sum_{j=1}^{\ModeNumber} \rounds{2 n_j + 1} } + \nonumber \\
  \PhaseSpaceDispAmp^2 \sum_{j,j'} \Big[ & n_{j'} \DensityMatrixElement_{n_{j},n_{j},n_{j'}-1,n_{j'}-1}^{n_1,\dots,n_\ModeNumber;{j,j'}}  +
 (n_{j'}+1) \DensityMatrixElement_{n_{j},n_{j},n_{j'}+1,n_{j'}+1}^{n_1,\dots,n_\ModeNumber;{j,j'}}
 +
 n_{j} \DensityMatrixElement_{n_{j}-1,n_{j}-1,n_{j'},n_{j'}}^{n_1,\dots,n_\ModeNumber;{j,j'}}  +
 (n_{j}+1) \DensityMatrixElement_{n_{j}+1,n_{j}+1,n_{j'},n_{j'}}^{n_1,\dots,n_\ModeNumber;{j,j'}}\Big] +\nonumber \\
  \PhaseSpaceDispAmp^2 e^{-\stdev_{\DispPhase}^2}\sum_{j,j'} \Big[ -& \sqrt{n_j} \sqrt{n_{j'}+1} \DensityMatrixElement_{n_{j}-1,n_{j},n_{j'}+1,n_{j'}}^{n_1,\dots,n_\ModeNumber;{j,j'}} + \sqrt{n_j} \sqrt{n_{j'}} \DensityMatrixElement_{n_{j}-1,n_{j},n_{j'},n_{j'}-1}^{n_1,\dots,n_\ModeNumber;{j,j'}} \nonumber \\ - &
 \sqrt{n_j+1} \sqrt{n_{j'}} \DensityMatrixElement_{n_{j}+1,n_{j},n_{j'}-1,n_{j'}}^{n_1,\dots,n_\ModeNumber;{j,j'}} + \sqrt{n_j+1} \sqrt{n_{j'}+1} \DensityMatrixElement_{n_{j}+1,n_{j},n_{j'},n_{j'}+1}^{n_1,\dots,n_\ModeNumber;{j,j'}}  + \nonumber \\
 & \sqrt{n_j} \sqrt{n_{j'}} \DensityMatrixElement_{n_{j},n_{j}-1,n_{j'}-1,n_{j'}}^{n_1,\dots,n_\ModeNumber;{j,j'}} - \sqrt{n_j} \sqrt{n_{j'}+1} \DensityMatrixElement_{n_{j},n_{j}-1,n_{j'},n_{j'}+1}^{n_1,\dots,n_\ModeNumber;{j,j'}} +\nonumber \\
 & 
 \sqrt{n_j+1} \sqrt{n_{j'}+1} \DensityMatrixElement_{n_{j},n_{j}+1,n_{j'}+1,n_{j'}}^{n_1,\dots,n_\ModeNumber;{j,j'}} - \sqrt{n_j+1} \sqrt{n_{j'}} \DensityMatrixElement_{n_{j},n_{j}+1,n_{j'},n_{j'}-1}^{n_1,\dots,n_\ModeNumber;{j,j'}} 
 \Big] + \mathcal{O}\rounds{\PhaseSpaceDispAmp^3}.
 \label{pnin2}
\end{align}
The Fisher information is then
\begin{align}
   \Fisher_\PhaseSpaceDispAmp = &4 \sum_{m_1,\dots,m_\ModeNumber}  \sum_{j} \DensityMatrixElement_{m_1,m_1,\dots,m_\ModeNumber,m_\ModeNumber} (2m_i+1)  + \nonumber \\ &4 e^{-\stdev_{\DispPhase}^2} \sum_{m_1,\dots,m_\ModeNumber}  \sum_{j\neq j'} \rounds{\sqrt{m_j+1} \sqrt{m_{j'}}\DensityMatrixElement^{m_1,\dots,m_\ModeNumber}_{\text{R};j,j'} + \sqrt{m_j} \sqrt{m_{j'}+1}\DensityMatrixElement^{m_1,\dots,m_\ModeNumber}_{\text{L};j,j'}}.
\end{align}
For the optimal states, it gives the following scaling,
\begin{align}
    \CFisher \rounds{\DensityMatrix_\DispLetter} = 4 \ModeNumber (2 \AverN+1)  + 8 e^{-\stdev_{\DispPhase}^2} \ModeNumber (\ModeNumber-1) \AverN .
\end{align}

\subsection{Loss and motional heating}
Let us analyze the effect of the loss and motional heating on the scheme.
We start with the master equation,
\begin{align}
    \partial_{\Time} \DensityMatrix (\Time) &=  \sum_{i,j} \frac{1}{2} \squares{2 \Coll_{ij} \DensityMatrix(\Time) \Coll_{ij}^{\dagger} - \DensityMatrix(\Time ) \Coll_{ij}^{\dagger} \Coll_{ij} - \Coll_{ij}^{\dagger} \Coll_{ij} \DensityMatrix(\Time )   }, \nonumber \\
    \Coll_{i1} &= \sqrt{\PhotonLossCoeff \rounds{1+\ThermalOccupation}} \Annihilation_i, \ \ \Coll_{i2} =  \sqrt{\PhotonLossCoeff \ThermalOccupation} \Creation_i,
    \label{full-master1}
\end{align}
where the limit $\ThermalOccupation \rightarrow 0$ constitutes the loss case, while the limit $\ThermalOccupation \gg 1 $ is the heating case.
We will additionally consider the limit of small decoherence, $\SmallTime = \PhotonLossCoeff \Time \ll 1$ and $\SmallTimeBar = \SmallTime \ThermalOccupation \ll 1$, for the loss and heating cases, respectively. 
It yields in the loss case,
\begin{align}
\DensityMatrix_\LossLetter =  \DensityMatrix - \SmallTime \sum_i \squares{ \frac{1}{2}  \rounds{\DensityMatrix \NumberOperator_i + \NumberOperator_i \DensityMatrix} -  \Annihilation_i \DensityMatrix \Creation_i},
\end{align}
and in the heating case,
\begin{align}
\DensityMatrix_\HeatLetter =  \DensityMatrix\rounds{1-\ModeNumber\SmallTimeBar} - \SmallTimeBar  \sum_i \squares{  \rounds{ \DensityMatrix \NumberOperator_i + \NumberOperator_i \DensityMatrix} - \rounds{   \Creation_i \DensityMatrix \Annihilation_i +  \Annihilation_i \DensityMatrix \Creation_i}}.
\end{align}

First, let us analyze which noise-affected terms change the final Fisher information in the leading order.
In both measurement schemes we consider, only the diagonal entries in the density matrix are measured.
In general, they can be affected in the linear order, such that
\begin{align}
    \Prob_\PhaseSpaceDispAmp = \SmallTime g_1 + \PhaseSpaceDispAmp^2 g_2 + \mathcal{O}(\SmallTime \PhaseSpaceDispAmp^2),
\end{align}
where it is already assumed that in the noiseless case, the leading order of the contribution is $\mathcal{O}(\PhaseSpaceDispAmp^2)$.
Then, the contribution to the total Fisher information is
\begin{align}
    \SmallFisher_\PhaseSpaceDispAmp = \frac{(\partial_\PhaseSpaceDispAmp \Prob_\PhaseSpaceDispAmp)^2}{\Prob_\PhaseSpaceDispAmp} = \frac{4 g_2}{1+\frac{\SmallTime}{\PhaseSpaceDispAmp^2} \frac{g_1}{g_2}} + \mathcal{O}(\SmallTime \PhaseSpaceDispAmp^2) = 4 g_2 - 4 \frac{\SmallTime}{\PhaseSpaceDispAmp^2} g_1 + \mathcal{O}(\SmallTime ) +  \mathcal{O}(\SmallTime \PhaseSpaceDispAmp^2) ,
\end{align}
where the last expression assumes that $\SmallTime / \PhaseSpaceDispAmp^2 \ll 1$.
Notably, corrections to $g_1$ and $g_2$ due to noise enter in a subleading way.
Note that condition $\SmallTime / \PhaseSpaceDispAmp^2 \ll 1$ implies that linear corrections in just $\SmallTime$ are smaller than corrections of order $\mathcal{O}(\PhaseSpaceDispAmp^2)$, hence we omit them.
As such, for the leading noise correction, we need to focus on the corrections to $\Prob_{\PhaseSpaceDispAmp}\rounds{n_1,\dots,n_\ModeNumber} $ linear in $\SmallTime$ or $\SmallTimeBar$.

We can explicitly write the effect of loss and heating on~\eqref{listing}, assuming the checkerboard states.
Let us start with the loss,
\begin{align}
\Prob_\PhaseSpaceDispAmp^{\LossLetter}\rounds{m_1,\dots,m_j+1,\dots,m_{j'}\dots,m_\ModeNumber}  &=  \PhaseSpaceDispAmp^2 (m_j+1) \DensityMatrixElement^{m_1,\dots,m_\ModeNumber}_{j,j'}  + \PhaseSpaceDispAmp^2 \sqrt{m_j+1} \sqrt{m_{j'}}\DensityMatrixElement^{m_1,\dots,m_\ModeNumber}_{\text{R};j,j'}, \nonumber \\
\Prob_\PhaseSpaceDispAmp^{\LossLetter}\rounds{m_1,\dots,m_j-1,\dots,m_{j'}\dots,m_\ModeNumber}  &=  \rounds{\SmallTime+\PhaseSpaceDispAmp^2}  m_j \DensityMatrixElement^{m_1,\dots,m_\ModeNumber}_{j,j'}  + \PhaseSpaceDispAmp^2 \sqrt{m_j} \sqrt{m_{j'}+1}\DensityMatrixElement^{m_1,\dots,m_\ModeNumber}_{\text{L};j,j'},
\nonumber \\
\Prob_\PhaseSpaceDispAmp^{\LossLetter}\rounds{m_1,\dots,m_j,\dots,m_{j'}+1\dots,m_\ModeNumber}  &= \PhaseSpaceDispAmp^2 (m_{j'}+1) \DensityMatrixElement^{m_1,\dots,m_\ModeNumber}_{j,j'}  + \PhaseSpaceDispAmp^2 \sqrt{m_j} \sqrt{m_{j'}+1}\DensityMatrixElement^{m_1,\dots,m_\ModeNumber}_{\text{L};j,j'},
\nonumber \\
\Prob_\PhaseSpaceDispAmp^{\LossLetter}\rounds{m_1,\dots,m_j,\dots,m_{j'}-1\dots,m_\ModeNumber}  &=  \rounds{\SmallTime+\PhaseSpaceDispAmp^2} m_{j'} \DensityMatrixElement^{m_1,\dots,m_\ModeNumber}_{j,j'}  + \PhaseSpaceDispAmp^2 \sqrt{m_j+1} \sqrt{m_{j'}}\DensityMatrixElement^{m_1,\dots,m_\ModeNumber}_{\text{R};j,j'},\label{listing3}
\end{align}
which implies the change in Fisher information,
\begin{align}
   \Fisher_\PhaseSpaceDispAmp = &4 \sum_{m_1,\dots,m_\ModeNumber}  \sum_{j} \DensityMatrixElement_{m_1,m_1,\dots,m_\ModeNumber,m_\ModeNumber} \squares{m_i+1 + m_i \rounds{1 - \frac{\SmallTime}{\PhaseSpaceDispAmp^2}}} + \nonumber \\ &4 \sum_{m_1,\dots,m_\ModeNumber}  \sum_{j\neq j'} \rounds{\sqrt{m_j+1} \sqrt{m_{j'}}\DensityMatrixElement^{m_1,\dots,m_\ModeNumber}_{\text{R};j,j'} + \sqrt{m_j} \sqrt{m_{j'}+1}\DensityMatrixElement^{m_1,\dots,m_\ModeNumber}_{\text{L};j,j'}}.
\end{align}
Similarly, for the heating, we have
\begin{align}
\Prob_\PhaseSpaceDispAmp^{\HeatLetter}\rounds{m_1,\dots,m_j+1,\dots,m_{j'}\dots,m_\ModeNumber}  &=  \rounds{\SmallTimeBar+\PhaseSpaceDispAmp^2}  (m_j+1) \DensityMatrixElement^{m_1,\dots,m_\ModeNumber}_{j,j'}  + \PhaseSpaceDispAmp^2 \sqrt{m_j+1} \sqrt{m_{j'}}\DensityMatrixElement^{m_1,\dots,m_\ModeNumber}_{\text{R};j,j'}, \nonumber \\
\Prob_\PhaseSpaceDispAmp^{\HeatLetter}\rounds{m_1,\dots,m_j-1,\dots,m_{j'}\dots,m_\ModeNumber}  &=  \rounds{\SmallTimeBar+\PhaseSpaceDispAmp^2}  m_j \DensityMatrixElement^{m_1,\dots,m_\ModeNumber}_{j,j'}  + \PhaseSpaceDispAmp^2 \sqrt{m_j} \sqrt{m_{j'}+1}\DensityMatrixElement^{m_1,\dots,m_\ModeNumber}_{\text{L};j,j'},
\nonumber \\
\Prob_\PhaseSpaceDispAmp^{\HeatLetter}\rounds{m_1,\dots,m_j,\dots,m_{j'}+1\dots,m_\ModeNumber}  &= \rounds{\SmallTimeBar+\PhaseSpaceDispAmp^2}  (m_{j'}+1) \DensityMatrixElement^{m_1,\dots,m_\ModeNumber}_{j,j'}  + \PhaseSpaceDispAmp^2 \sqrt{m_j} \sqrt{m_{j'}+1}\DensityMatrixElement^{m_1,\dots,m_\ModeNumber}_{\text{L};j,j'},
\nonumber \\
\Prob_\PhaseSpaceDispAmp^{\HeatLetter}\rounds{m_1,\dots,m_j,\dots,m_{j'}-1\dots,m_\ModeNumber}  &=  \rounds{\SmallTimeBar+\PhaseSpaceDispAmp^2} m_{j'} \DensityMatrixElement^{m_1,\dots,m_\ModeNumber}_{j,j'}  + \PhaseSpaceDispAmp^2 \sqrt{m_j+1} \sqrt{m_{j'}}\DensityMatrixElement^{m_1,\dots,m_\ModeNumber}_{\text{R};j,j'},\label{listing5}
\end{align}
and again the change in Fisher information,
\begin{align}
   \Fisher_\PhaseSpaceDispAmp = &4 \sum_{m_1,\dots,m_\ModeNumber}  \sum_{j} \DensityMatrixElement_{m_1,m_1,\dots,m_\ModeNumber,m_\ModeNumber} \rounds{2 m_i +1}\rounds{1 - \frac{\SmallTimeBar}{\PhaseSpaceDispAmp^2}} + \nonumber \\ &4 \sum_{m_1,\dots,m_\ModeNumber}  \sum_{j\neq j'} \rounds{\sqrt{m_j+1} \sqrt{m_{j'}}\DensityMatrixElement^{m_1,\dots,m_\ModeNumber}_{\text{R};j,j'} + \sqrt{m_j} \sqrt{m_{j'}+1}\DensityMatrixElement^{m_1,\dots,m_\ModeNumber}_{\text{L};j,j'}}.
\end{align}

\subsection{Dephasing}
Now, let us turn our attention to the effect of dephasing.
The relevant entries in the density matrix for the considered scheme are the ones from Eq.~\eqref{entries}.
Due to the dephasing given by the Lindblad equation,
\begin{align}
    \partial_{\Time} \DensityMatrix (\Time) &=  \sum_{i} \frac{1}{2} \squares{2 \Coll_{i} \DensityMatrix(\Time) \Coll_{i}^{\dagger} - \DensityMatrix(\Time ) \Coll_{i}^{\dagger} \Coll_{i} - \Coll_{i}^{\dagger} \Coll_{i} \DensityMatrix(\Time )   }, \nonumber \\
    \Coll_{i} &= \sqrt{\DephasingCoeff} \Creation_i \Annihilation_i, 
    \label{full-master2}
\end{align}
these entries are affected in the following way:
\begin{align}
\DensityMatrixElement^{m_1,\dots,m_\ModeNumber}_{j,j'} &\rightarrow  \DensityMatrixElement^{m_1,\dots,m_\ModeNumber}_{j,j'}, \nonumber \\
\DensityMatrixElement^{m_1,\dots,m_\ModeNumber}_{\text{R};j,j'} &\rightarrow \DensityMatrixElement^{m_1,\dots,m_\ModeNumber}_{\text{R};j,j'}  e^{-\DephasingCoeff \Time } , \nonumber \\
\DensityMatrixElement^{m_1,\dots,m_\ModeNumber}_{\text{L};j,j'}  &\rightarrow \DensityMatrixElement^{m_1,\dots,m_\ModeNumber}_{\text{L};j,j'} e^{-\DephasingCoeff \Time }.
\end{align}
The Fisher information is then
\begin{align}
   \Fisher_\PhaseSpaceDispAmp = &4 \sum_{m_1,\dots,m_\ModeNumber}  \sum_{j} \DensityMatrixElement_{m_1,m_1,\dots,m_\ModeNumber,m_\ModeNumber} (2m_i+1) + \nonumber \\ &4 e^{-\DephasingCoeff \Time } \sum_{m_1,\dots,m_\ModeNumber}  \sum_{j\neq j'} \rounds{\sqrt{m_j+1} \sqrt{m_{j'}}\DensityMatrixElement^{m_1,\dots,m_\ModeNumber}_{\text{R};j,j'} + \sqrt{m_j} \sqrt{m_{j'}+1}\DensityMatrixElement^{m_1,\dots,m_\ModeNumber}_{\text{L};j,j'}}.
\end{align}

\newpage

\section{Two-mode examples}
\begin{figure}[ht!]
    \includegraphics[width=\linewidth]{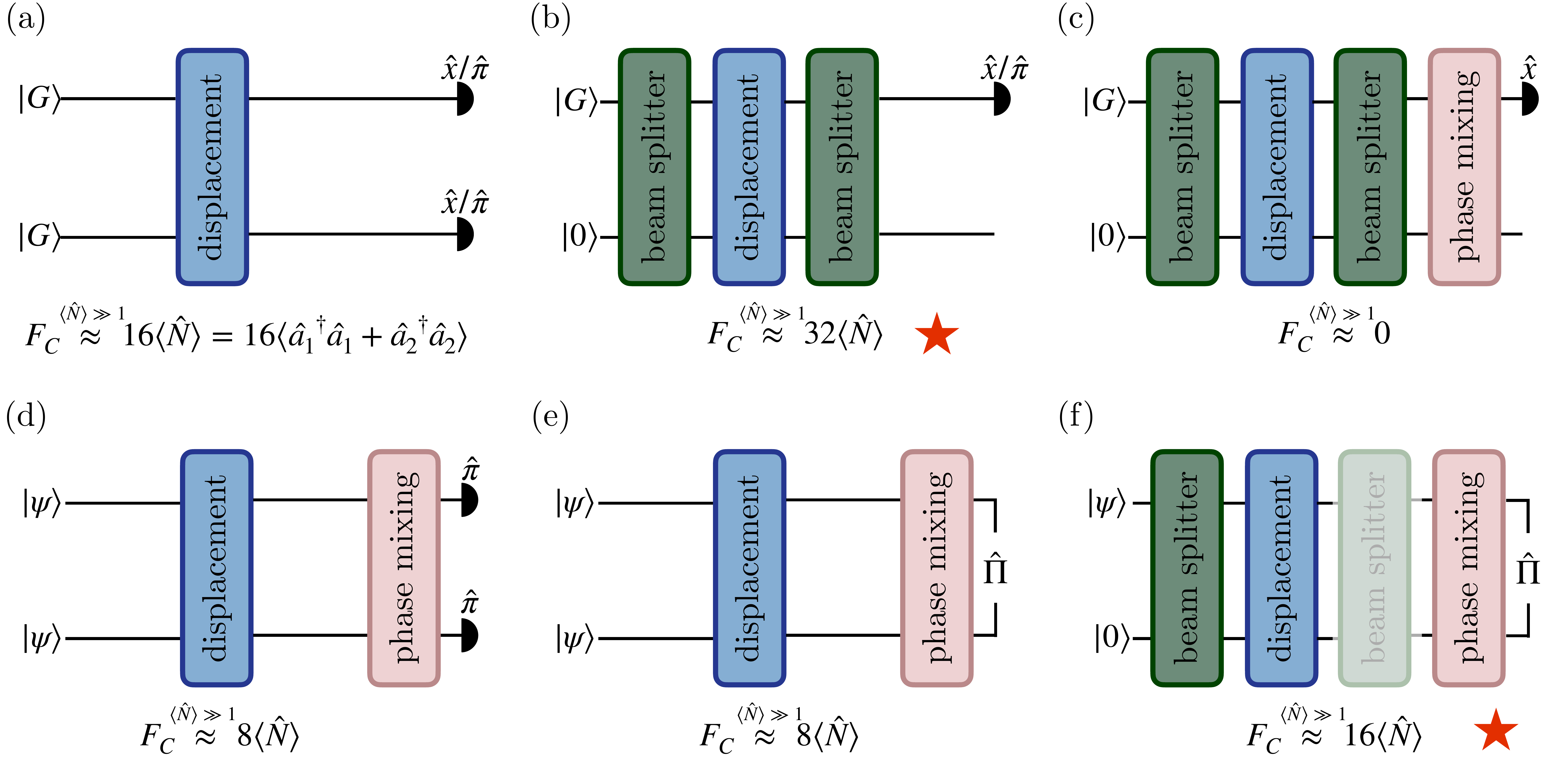}  
    \caption{The phase-sensitive and phase-insensitive measurement schemes considered in this comparison. If there is no phase mixing involved, it is assumed that the displacement direction is known and aligned optimally with respect to the probe state and the homodyne detection. $\PosOp$ signifies homodyne detection, $\hat{\pi}$ single-mode parity measurement, while $\hat{\Pi} = \hat{\pi}_1 \hat{\pi}_2$ is the joint parity measurement. The box with the label displacement means equal phase-space displacement in the modes; the beam-splitter label means a balanced beam-splitter; however, if it is the second in the order, it is the reverse of the first one; the 'phase mixing' box signifies that the phase of the displacement varies from shot to shot. State $\ket{G}$ is the squeezed vacuum (ground state), and $\ket{\psi}$ is an arbitrary state with a definite parity (that also includes the squeezed vacuum). The top row (a-c) describes the usual homodyne scheme for the force sensing, while the bottom one (d-f) concerns the schemes described in the main text involving parity measurements. (a) Two modes are prepared in the same squeezed vacuum state and then, after the action of displacement, are independently measured via a homodyne or the parity measurement is made. Such a scheme gives twice the quantum Fisher information of a single-mode scheme, which in the large-$\langle \FullNOp \rangle$ and small-$\PhaseSpaceDispAmp$ limit gives $\CFisher \approx 16 \langle \FullNOp \rangle$ [cf. Eqs.~\eqref{eqS1a} and~\eqref{GaussPar}], where $\FullNOp = \Creation_1 \Annihilation_1 + \Creation_2 \Annihilation_2$. (b) The usual interferometric scheme, where a Gaussian state is prepared in one of the modes, then split into two modes via a beam splitter, recombined back into the first mode, and then a homodyne or parity measurement is made. Such a scheme gives the same Fisher information as a scheme in (a) if the same state $\ket{G}$ is used; however, the total number of excitations $\langle \FullNOp \rangle$ is lower, hence the saturation of the bound (24) from the main text at $\CFisher \approx 32 \langle \FullNOp \rangle$ [see Eqs.~\eqref{eqS1b} and~\eqref{GaussPar}]. This saturation is signified by the red star. (c) The same scheme as in (b), however, in this case, the phase is not fixed between the shots. The Fisher information tends to 0 as $\PhaseSpaceDispAmp \rightarrow 0$ [cf. Eq.~\eqref{eqS1c}], showing inadequacy of this usual scheme in the phase-scrambled case. (d) The 'single-mode' case, where a state $\ket{\psi}$ is prepared independently in both modes, and then each of the modes undergoes the parity measurement. In this case, the phase is also randomized. The Fisher information is lower than in the analogous case (a); however, it also scales with $\langle \FullNOp \rangle$. Homodyne measurement in this case would yield no advantage. (e) The same scheme as in (d), but this time, the total parity is measured [a binary observable, unlike in case (d)]. The Fisher information is the same as in the case (d). (f) The scheme analogous to (b)---a single mode is prepared in the state $\ket{\psi}$, and then split into two modes. As the measurement is the total parity, the recombination by the second beam-splitter is optional as passive operations preserve the parity. However, such a recombination allows for performing the total parity measurement by only a single local parity detector, bringing the amount of the needed resources to the same level as (b). Similarly to (b), this scheme saturates the theoretical bound given by Eq. (10) in the main text and scales with $\langle \FullNOp \rangle$ even with phase scrambling.
  \label{FigS0}}
\end{figure}

Here we consider in detail the $\ModeNumber = 2$ case, comparing the usual scheme involving Gaussian states and homodyne detection with the two schemes involving checkerboard states and parity measurements that are considered in the main text.
The detailed description of the considered cases is depicted in Fig.~\ref{FigS0}.

The starting point for the comparison is the strategies used in the force sensing when the direction of the phase-space displacement is known and fixed, such that the state and the measurement can be prepared adequately.
These strategies commonly utilize Gaussian states and homodyne measurement.
The classical Fisher information with respect to some parameter $\SenParam$ for the homodyne measurement of quadrature $\PosOp_\QAngle$ reads
\begin{align}
    \CFisher^{\text{hom}} = \int_{-\infty}^{\infty} \dd x \frac{[{\partial_{\SenParam} \Prob_{\SenParam} (x)}]^2}{\Prob_{\SenParam} (x)},
\end{align}
which for Gaussian states with mean $\mu_\QAngle$ and variance $\sigma_\QAngle$ reads
\begin{align}
    \CFisher^{\text{hom}} = \frac{(\partial_\SenParam \mu_\QAngle)^2}{\sigma_\QAngle} + \frac{1}{2} \frac{(\partial_\SenParam \sigma_\QAngle)^2}{\sigma_\QAngle^2}.
\end{align}
If we take a squeezed vacuum state with a squeezing parameter $\Squeeze$ as a probe, and we aim to sense displacement $\PhaseSpaceDispAmp$, then the variance is not affected by the displacement, but $\mu_\QAngle = \sqrt{2} \PhaseSpaceDispAmp \cos \QAngle$, giving
\begin{align}
    \CFisher^{\text{hom}} = \frac{4 \cos^2 \QAngle}{e^{-2 \Squeeze} \cos^2 \QAngle + e^{2 \Squeeze} \sin^2 \QAngle}.
\end{align}
Then, choosing optimal $\QAngle = 0$, we get
\begin{align}
    \CFisher^{\text{hom}} = 4 e^{2 \Squeeze}.
\end{align}
Let us now then consider three different schemes.
The first one is depicted in Fig.~\ref{FigS0}(a).
Each of the two modes is prepared in the same squeezed vacuum state and then measured by homodyne.
Then, the classical Fisher information is twice the optimal single-mode value,
\begin{align}
    \CFisher^{\text{hom}} = 8 e^{2 \Squeeze} = 8 (2 \AvN + 1 + 2\sqrt{\AvN(\AvN+1)}) \approx 32 \AvN = 16 \langle \FullNOp \rangle,
    \label{eqS1a}
\end{align}
where $\AvN$ is the average number of excitations in a single-mode input state and $\FullNOp = \Creation_1 \Annihilation_1 + \Creation_2 \Annihilation_2$.
The second scheme [presented in Fig.~\ref{FigS0}(b)] involves preparation of a squeezed vacuum state in one of the modes, splitting it by a balanced beam-splitter, recombining it back, and homodyne measurement on the originally occupied mode.
In $\ModeNumber=2$, the following relation holds,
\begin{align}
\BogoUnitary^\dagger[\hat{D}_1(\CohDis)\hat{D}_2(\CohDis)]\BogoUnitary = \hat{D}_1(\sqrt{2}\CohDis)\hat{D}_2(0),
\label{ident}
\end{align}
which means that the scheme simplifies a single-mode sensing with the displacement amplitude enhanced by a factor $\sqrt{2}$.
Then, it means that the Fisher information is just
\begin{align}
    \CFisher^{\text{hom}} = 8 e^{2 \Squeeze} = 8 (2 \AvN + 1 + 2\sqrt{\AvN(\AvN+1)}) \approx 32 \AvN = 32 \langle \FullNOp \rangle,
    \label{eqS1b}
\end{align}
where this time $\langle \FullNOp \rangle = \AvN $, saturating quantum Fisher information~\cite{Kwon2022}.
Despite the same values in Eq.~\eqref{eqS1a} and Eq.~\eqref{eqS1b}, the resources needed for the generation of these states are different---in the latter case, one needs to prepare a specific state only in a single mode, while in the former one, in both.
The last case involves the same scheme as the second one, but with the phase mixing considered in the main text.
Then, the displacement of the state is not affected, only the variance,
\begin{align}
    \sigma = \frac{1}{2} e^{-2 \Squeeze} + \PhaseSpaceDispAmp^2,
\end{align}
where the multiplicative factor $\sqrt{2}$ due to two-mode scheme has already been accounted for.
Then, the Fisher information is
\begin{align}
    \CFisher^{\text{hom}} = \frac{2 \PhaseSpaceDispAmp^2}{(\frac{1}{2} e^{-2 \Squeeze} + \PhaseSpaceDispAmp)^2} \approx 128 \PhaseSpaceDispAmp^2 \AvN^2,
    \label{eqS1c}
\end{align}
where the rightmost expression is taken in the limit of large $\AvN$ and small $\PhaseSpaceDispAmp$.
Notably, all the advantage is lost as $\CFisher^{\text{hom}} \rightarrow 0$ when $\PhaseSpaceDispAmp \rightarrow 0$.

\begin{figure}[ht!]
    \includegraphics[width=0.55\linewidth]{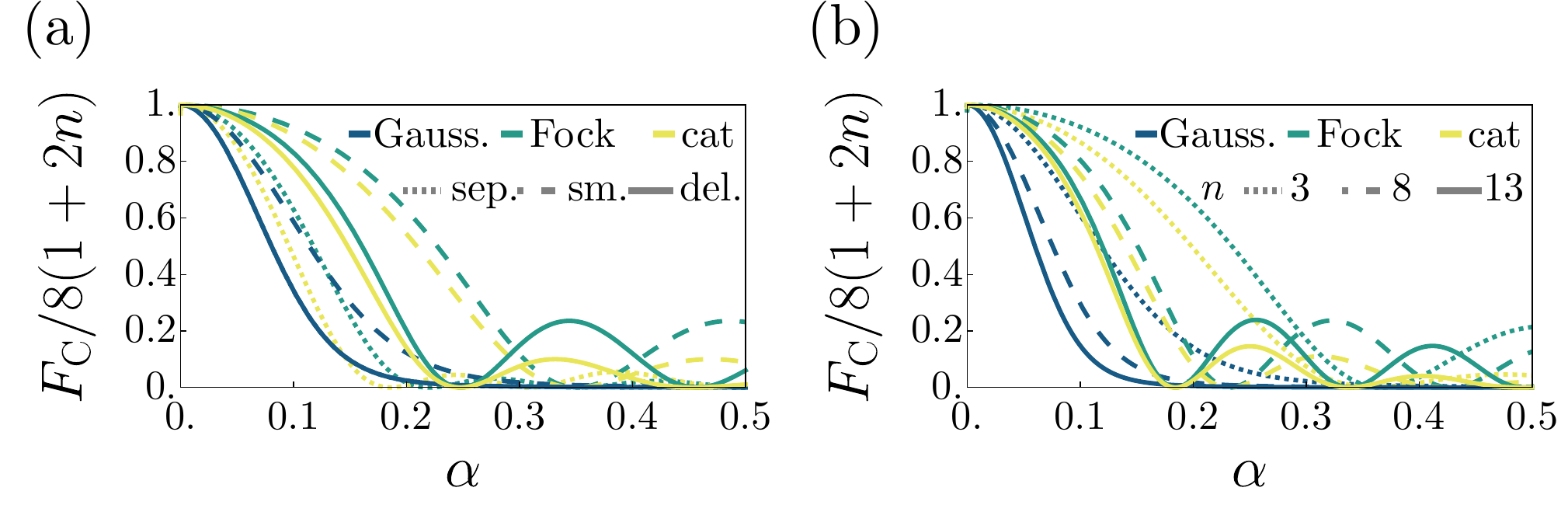}  
    \caption{Fisher information of parity measurement for different two-mode states as a function of measured displacement $\PhaseSpaceDispAmp$. Both plots provide a comparison between single-mode squeezed vacuum (blue, Gauss.), Fock (green), and cat (yellow) states that are prepared in both modes independently, and then a joint-parity measurement is taken (dotted, 'separable' case, sep.), are prepared in both modes independently, and then local parity measurements are taken without computing total parity (dashed, 'single-mode' case, sm.), are prepared in a single mode and then delocalized via a beam-splitter (solid, 'delocalized' case, del.). (a) Comparison of the states with the same average phonon number $n=5$. In every case, Gaussian states exhibit a smaller dynamical range (range of $\PhaseSpaceDispAmp$ such that $\Fisher>8 \ModeNumber$) than cat and Fock states, in that order. Single-mode strategy exhibits the largest dynamical range, followed by the delocalized and separable ones. This is contrary to the phase-fixed case, where both single-mode and delocalized strategies exhibit infinite dynamical ranges in the noiseless scenario. (b) Comparison of states with different average occupation numbers for the delocalized strategy. The larger the occupation number, the faster the metrological advantage decays as a function of $\PhaseSpaceDispAmp$.
  \label{FigS1}}
\end{figure}

Let us now switch to how to circumvent the problem of such a scaling in the phase-mixed case, namely, to the two strategies presented in the main text,
\begin{align}
    \DensityMatrix_{\text{bs}} &= \BogoUnitary \DensityMatrix_{\WaveFunctionC} \otimes \DensityMatrix_{0} \BogoUnitary^\dagger, \nonumber \\
    \DensityMatrix_{\text{sep}} &= \DensityMatrix_{\WaveFunctionC} \otimes  \DensityMatrix_{\WaveFunctionC} ,
\end{align}
with
\begin{align}
    \BogoUnitary^\dagger \Annihilation_1 \BogoUnitary = \frac{\Annihilation_1 + \Annihilation_2}{\sqrt{2}}, \quad \BogoUnitary^\dagger \Annihilation_2 \BogoUnitary = \frac{\Annihilation_1 - \Annihilation_2}{\sqrt{2}}, \quad \DensityMatrix_{\WaveFunctionC} = \ket{\WaveFunctionC}\bra{\WaveFunctionC}, \quad \DensityMatrix_{0} = \ket{0}\bra{0},
\end{align}
where the first scheme is dubbed 'delocalized' and the other 'separable'.
Following identity~\eqref{ident}, we can write for $\DensityMatrix_{\text{bs}}$,
\begin{align}
     \langle \ParityOp \rangle_{\text{bs}} &= \int_0^{2 \pi} \frac{\dd \DispPhase}{2 \pi} \Tr[  \hat{D}_2(\PhaseSpaceDispAmp e^{\ImagUnit \DispPhase} ) \hat{D}_1(\PhaseSpaceDispAmp e^{\ImagUnit \DispPhase}) \BogoUnitary \DensityMatrix_{\WaveFunctionC} \otimes \DensityMatrix_{0} \BogoUnitary^\dagger \hat{D}_1^\dagger(\PhaseSpaceDispAmp e^{\ImagUnit \DispPhase} )\hat{D}_2^\dagger(\PhaseSpaceDispAmp e^{\ImagUnit \DispPhase} )  \ParityOp] \nonumber \\
     &=\int_0^{2 \pi} \frac{\dd \DispPhase}{2 \pi} \Tr[ \hat{D}_1(\sqrt{2}\PhaseSpaceDispAmp e^{\ImagUnit \DispPhase}) \DensityMatrix_{\WaveFunctionC} \otimes \DensityMatrix_{0} \hat{D}_1^\dagger(\sqrt{2}\PhaseSpaceDispAmp e^{\ImagUnit \DispPhase})  \ParityOp] \nonumber \\
     &=\ParityValue \int_0^{2 \pi} \frac{\dd \DispPhase}{2 \pi} \Tr[ \hat{D}_1(-2\sqrt{2}\PhaseSpaceDispAmp e^{\ImagUnit \DispPhase}) \DensityMatrix_{\WaveFunctionC}] =\ParityValue \int_0^{2 \pi} \frac{\dd \DispPhase}{2 \pi} \Characteristic_{\WaveFunctionC}(-2\sqrt{2}\PhaseSpaceDispAmp e^{\ImagUnit \DispPhase}), 
\end{align}
while for the separable state,
\begin{align}
\langle \ParityOp \rangle_{\text{sep}} = \int_0^{2 \pi} \frac{\dd \DispPhase}{2 \pi} \squares{\Characteristic_{\WaveFunctionC}(-2\PhaseSpaceDispAmp e^{\ImagUnit \DispPhase})}^2.
\end{align}
It is also worth comparing with the 'single-mode' strategy, where one retrieves information about parity in every mode.
For the case with a beam splitter, nothing changes, as all the information is in a single mode.
Hence, the Fisher information can be computed as
\begin{align}
    \Fisher_\text{bs} = \CFisher (\Measurement^{\Parity},\DensityMatrix_\text{bs}) = \frac{(\partial_\PhaseSpaceDispAmp \langle \ParityOp \rangle_{\text{bs}})^2}{1 - \langle \ParityOp \rangle_{\text{bs}}^2},
    \label{Fbs}
\end{align}
However, for a separable state, there is a difference.
Effectively, we perform two measurements of two commuting observables $\ParityOp_1$ and $\ParityOp_2$, each giving
\begin{align}
\langle \ParityOp \rangle_{1} = \int_0^{2 \pi} \frac{\dd \DispPhase}{2 \pi} \Characteristic_{\WaveFunctionC}(-2\PhaseSpaceDispAmp e^{\ImagUnit \DispPhase})
\end{align}
with associated Fisher information
\begin{align}
    \Fisher_\text{sm} = 2\CFisher (\Measurement^{\Parity_1},\DensityMatrix_\text{sep}) = 2\frac{(\partial_\PhaseSpaceDispAmp \langle \ParityOp \rangle_{1})^2}{1 - \langle \ParityOp \rangle_{1}^2},
    \label{Fsm}
\end{align}
where the factor 2 comes from the fact that we have two copies of the state $\DensityMatrix_{\WaveFunctionC}$.
On the other hand, if we perform a joint parity measurement (or, in other words, we erase information about the parity of distinctive modes), we have
\begin{align}
    \Fisher_\text{sep} = \CFisher (\Measurement^{\Parity},\DensityMatrix_\text{sep}) = \frac{(\partial_\PhaseSpaceDispAmp \langle \ParityOp \rangle_{\text{sep}})^2}{1 - \langle \ParityOp \rangle_{\text{sep}}^2}.
    \label{Fsep}
\end{align}
Let us consider then the three families of single-mode states with definite parity, Fock, squeezed vacuum, and cat states,
\begin{align}
    \Characteristic_\text{Fock} (\beta,\DispPhase) &= e^{-\beta^2/2} L_n (\beta^2), \nonumber \\
    \Characteristic_\text{sq} (\beta,\DispPhase) &= e^{-\frac{\alpha^2}{2} (e^{-2r} \cos^2 \DispPhase + e^{2r} \sin^2 \DispPhase  ) }, \nonumber \\
    \Characteristic_\text{cat} (\beta,\DispPhase) &= \frac{e^{-\beta^2/2}}{1+e^{-2 \gamma^2}} [ \cos(2 \gamma \beta \sin \DispPhase) + e^{-2 \gamma^2} \cosh(2 \gamma \beta \cos \DispPhase) ],
\end{align}
with $n = \sinh^2 r$ and $n = \gamma^2 \tanh \gamma^2$, and $n$ is average number of excitations in each case.
Let us introduce averaged characteristic functions,
\begin{align}
\overline{\Characteristic}_1 (\beta) &= \int_0^{2 \pi} \frac{\dd \DispPhase}{2 \pi} \Characteristic(\beta,\DispPhase), \nonumber \\
\overline{\Characteristic}_2 (\beta) &= \int_0^{2 \pi} \frac{\dd \DispPhase}{2 \pi} [\Characteristic(\beta,\DispPhase)]^2.
\label{mixinchis}
\end{align}
Then, we have
\begin{align}
\overline{\Characteristic}_{1,\text{Fock}} (\beta) &= e^{-\beta^2/2} L_n (\beta^2), \nonumber \\
\overline{\Characteristic}_{1,\text{sq}} (\beta) &= e^{-\frac{\beta^2(2n+1)}{2}} I_0\rounds{\beta^2 \sqrt{n(n+1)}}, \nonumber \\
\overline{\Characteristic}_{1,\text{cat}} (\beta) &= \frac{e^{-\beta^2/2}}{1+e^{-2 \gamma^2}} [ J_0(2 \gamma \beta ) + e^{-2 \gamma^2} I_0(2 \gamma \beta) ], \nonumber \\
\overline{\Characteristic}_{2,\text{Fock}} (\beta) &= e^{-\beta^2} [L_n (\beta^2)]^2, \nonumber \\
\overline{\Characteristic}_{2,\text{sq}} (\beta) &= e^{-\beta^2(2n+1)} I_0\rounds{2 \beta^2 \sqrt{n(n+1)}},\nonumber \\
\overline{\Characteristic}_{2,\text{cat}} (\beta) &= \frac{e^{-\beta^2}}{2\rounds{1+e^{-2 \gamma^2}}^2} [5+ J_0(4 \gamma \beta ) + e^{-4 \gamma^2}(1+ I_0(4 \gamma \beta)) ].
\end{align}
By plugging directly into Eqs.~\eqref{Fbs},~\eqref{Fsm},~\eqref{Fsep}, we get Fisher information as a function of $\DispPhase$, beyond the small-$\PhaseSpaceDispAmp$ expansion.
We plot the results in Fig.~\ref{FigS1}, where we provide their analysis in the caption.
On the other hand, if we do not perform the phase mixing in Eqs.~\eqref{mixinchis},
\begin{align}
\Characteristic_1 (\beta) &=  \Characteristic(\beta,\DispPhase), \nonumber \\
\Characteristic_2 (\beta) &= [\Characteristic(\beta,\DispPhase)]^2,
\label{mixinchis2}
\end{align}
we recover the phase-fixed case, but with the parity measurement instead of the homodyne one.
Explicit evaluation of the FI [Eqs.~\eqref{Fbs},~\eqref{Fsm}, and~\eqref{Fsep}] in the small-$\PhaseSpaceDispAmp$ limit then gives
\begin{align}
    \Fisher_\text{bs}^{\text{p-f}} (\ket{n}\ket{n}) &= \Fisher_\text{sm}^{\text{p-f}} (\ket{n}\ket{n})= \Fisher_\text{sep}^{\text{p-f}} (\ket{n}\ket{n}) = 8 (2 \AvN + 1 ), \\
    \Fisher_\text{bs}^{\text{p-f}} (\ket{G}\ket{G}) &= \Fisher_\text{sm}^{\text{p-f}} (\ket{G}\ket{G})= \Fisher_\text{sep}^{\text{p-f}} (\ket{G}\ket{G}) = 8 (2 \AvN + 1 + 2\sqrt{\AvN(\AvN+1)}), \label{GaussPar}  \\
     \Fisher_\text{bs}^{\text{p-f}} (\ket{\text{cat}}\ket{\text{cat}}) &= \Fisher_\text{sm}^{\text{p-f}} (\ket{\text{cat}}\ket{\text{cat}})= \Fisher_\text{sep}^{\text{p-f}} (\ket{\text{cat}}\ket{\text{cat}}) = 8 (2 \AvN + 1 ),
\end{align}
where the equalities are up to order $\mathcal{O}(\PhaseSpaceDispAmp^2)$.
Notably, parity measurement recovers the result for homodyne measurement with Gaussian states that is optimal in that case.

\clearpage
\newpage

\section{Derivation from the quantum Fisher information matrix}

In this section, we provide an alternative derivation of the bounds from the main text, with a more general starting point, involving different displacement amplitudes in different modes.
Such an approach allows for considering a broader range of physical scenarios.
We consider $M$ bosonic modes with annihilation operators $[\hat a_i,\hat a_j^\dagger] = \delta_{ij}$, where $i,j\in\{1,\dots,M\}$.
We introduce the generalized quadrature operator $\hat A_i(\phi) = (e^{i\phi} \hat a_i^\dagger - e^{-i\phi} \hat a_i)/i$,
and define the single-mode displacement of real amplitude $\alpha_i\in\mathbb{R}$ and phase $\phi$ as $\hat D_i(\alpha_i,\phi)
    = \exp\bigl(i \alpha_i \hat A_i(\phi)\bigr)$.
The multi-mode displacement with possibly different amplitudes $\boldsymbol{\alpha} = (\alpha_1,\dots,\alpha_M)^T$ but a common phase $\phi$ is
\begin{equation}
    \hat D(\boldsymbol{\alpha},\phi)
    = \prod_{i=1}^M \hat D_i(\alpha_i,\phi)
    = \exp\left( i \sum_{i=1}^M \alpha_i \hat A_i(\phi)\right).
\end{equation}

Let $\hat\rho$ be the probe state prepared at the input. The phase $\phi$ is uncontrollable and uniformly random in $[0,2\pi)$ in each experimental shot. The resulting phase-averaged output state is
\begin{equation}
    \hat\rho_{\boldsymbol{\alpha}}
    =
    \int_0^{2\pi}\frac{d\phi}{2\pi}\,
    \hat D(\boldsymbol{\alpha},\phi)\,\hat\rho\,\hat D^\dagger(\boldsymbol{\alpha},\phi).
    \label{eq:phase-avg-state}
\end{equation}
The quantum Fisher information matrix (QFIM) associated to the estimation of $\boldsymbol{\alpha}$ is defined by the symmetric logarithmic derivatives (SLDs) $\hat L_j$ satisfying
\begin{equation}
    \partial_{\alpha_j}\hat\rho_{\boldsymbol{\alpha}}
    = \frac{1}{2}\left(
      \hat L_j \hat\rho_{\boldsymbol{\alpha}}
      + \hat\rho_{\boldsymbol{\alpha}} \hat L_j
    \right),
\end{equation}
and
\begin{equation}
    \bigl[F_Q^{(\boldsymbol{\alpha})}(\hat\rho_{\boldsymbol{\alpha}})\bigr]_{ij}
    = \frac{1}{2}\Tr\left(
      \hat\rho_{\boldsymbol{\alpha}}
      \{\hat L_i,\hat L_j\}
    \right).
\end{equation}
The multi-parameter quantum Cramér--Rao bound (QCRB) states that for any unbiased estimator $\hat{\boldsymbol{\alpha}}$ based on $\nu$ independent measurement repetitions,
\begin{equation}
    \mathrm{Cov}(\hat{\boldsymbol{\alpha}})
    \succeq
    \frac{1}{\nu}\,
    \left[F_Q^{(\boldsymbol{\alpha})}(\hat\rho_{\boldsymbol{\alpha}})\right]^{-1},
    \label{eq:multi-QCRB}
\end{equation}
in the sense of matrix inequalities.

We can now compute this quantity for states that are pure before phase averaging. For a fixed $\phi$, define $\ket{\psi^{\phi}_{\boldsymbol{\alpha}}} := \hat D(\boldsymbol{\alpha},\phi)\ket{\psi}$
where $\ket{\psi}$ is a fixed pure probe state and $\hat\rho = \ket{\psi}\bra{\psi}$.
For a pure state $\ket{\psi^{\phi}_{\boldsymbol{\alpha}}}$ undergoing a unitary family generated by $\{\hat A_k(\phi)\}$, the QFIM has a particularly simple form:
\begin{equation}
    \bigl[F_Q^{(\boldsymbol{\alpha})}(\ket{\psi^\phi_{\boldsymbol{\alpha}}})\bigr]_{ij}
    =    4\,\mathrm{Cov}_{\ket{\psi^\phi_{\boldsymbol{\alpha}}}}\left(\hat A_i(\phi),\hat A_j(\phi)\right),
    \label{eq:pure-QFIM}
\end{equation}
where $\mathrm{Cov}_{\ket{\psi}}(\hat X,\hat Y) := \frac{1}{2}\expval{\hat X\hat Y + \hat Y \hat X}{\psi}
    - \expval{\hat X}{\psi}\expval{\hat Y}{\psi}$.
Because $\hat A_k(\phi)$ commutes with the displacement operator,
\begin{equation}
    \hat A_k(\phi)\ket{\psi^\phi_{\boldsymbol{\alpha}}} = \hat A_k(\phi)\hat D(\boldsymbol{\alpha},\phi)\ket{\psi}
    =
    \hat D(\boldsymbol{\alpha},\phi)\hat A_k(\phi)\ket{\psi},
\end{equation}
and the covariance is thus independent of $\boldsymbol{\alpha}$.
In particular, $\bigl[F_Q^{(\boldsymbol{\alpha})}(\ket{\psi^\phi_{\boldsymbol{\alpha}}})\bigr]_{ij} = 4\,\mathrm{Cov}_{\ket{\psi}}(\hat A_i(\phi),\hat A_j(\phi))$.

Given a QFIM for the vector $\boldsymbol{\alpha}$, the QFI for any \emph{single} parameter constructed from the linear combination
\begin{equation}
    \beta = \mathbf{u}^T \boldsymbol{\alpha}, \qquad \mathbf{u}\in\mathbb{R}^M,
\end{equation}
is obtained via error propagation as
\begin{equation}
    F_Q^{(\beta)}(\hat\rho_{\boldsymbol{\alpha}})
    =
    \left(\frac{\partial \boldsymbol{\alpha}}{\partial \beta}\right)^T
    F_Q^{(\boldsymbol{\alpha})}(\hat\rho_{\boldsymbol{\alpha}})
    \left(\frac{\partial \boldsymbol{\alpha}}{\partial \beta}\right).
\end{equation}
If $\boldsymbol{\alpha} = \beta\,\mathbf{u}$, then $\partial\boldsymbol{\alpha}/\partial \beta = \mathbf{u}$ and
\begin{equation}
    F_Q^{(\beta)}(\hat\rho_{\boldsymbol{\alpha}})
    =
    \mathbf{u}^T
    F_Q^{(\boldsymbol{\alpha})}(\hat\rho_{\boldsymbol{\alpha}})
    \mathbf{u}.
    \label{eq:scalar-from-matrix}
\end{equation}

In the main text, we are interested in estimating a \emph{global} force amplitude $\alpha$ applied identically to all modes, meaning
\begin{equation}
    \boldsymbol{\alpha} = \alpha\,\mathbf{1},
    \qquad
    \alpha \in \mathbb{R},
    \quad \mathbf{1}:=(1,\dots,1)^T,
\end{equation}
so from \eqref{eq:scalar-from-matrix} we have
\begin{equation}
    F_Q^{(\alpha)}(\hat\rho_{\alpha})
    =
    \mathbf{1}^T
    F_Q^{(\boldsymbol{\alpha})}(\hat\rho_{\boldsymbol{\alpha}})
    \mathbf{1}.
    \label{eq:FQ-alpha-as-quadratic-form}
\end{equation}
Thus, the scalar QFI for the global amplitude is the quadratic form of the multi-parameter QFIM along the direction specified by the all-ones vector.
For the pure-state family at fixed $\phi$, we can combine \eqref{eq:pure-QFIM} and \eqref{eq:FQ-alpha-as-quadratic-form} to obtain an expression in terms of variance of a collective generator. Define this collective generator as
\begin{equation}
    \hat G(\phi) := \sum_{i=1}^M \hat A_i(\phi).
\end{equation}
Then
\begin{align}
    \frac{1}{4}\,
    \mathbf{1}^T
    F_Q^{(\boldsymbol{\alpha})}\bigl(\ket{\psi_{\boldsymbol{\alpha}}^\phi}\bigr)
    \mathbf{1}
    &=
    \sum_{i,j=1}^M
    \mathrm{Cov}_{\ket{\psi}}(\hat A_i(\phi),\hat A_j(\phi))
\nonumber\\
    &= 
    \mathrm{Var}_{\ket{\psi}}\left(\sum_{i=1}^M \hat A_i(\phi)\right)
\nonumber\\
    &= \Delta^2_{\ket{\psi}} \hat G(\phi),
\end{align}
from which
\begin{equation}
    F_Q^{(\alpha)}\bigl(\ket{\psi_{\boldsymbol{\alpha}}^\phi}\bigr)
    =
    \mathbf{1}^T
    F_Q^{(\boldsymbol{\alpha})}\bigl(\ket{\psi_{\boldsymbol{\alpha}}^\phi}\bigr)
    \mathbf{1}
    =
    4\,\Delta^2_{\ket{\psi}} \hat G(\phi).
    \label{eq:scalar-QFI-pure}
\end{equation}
We will upper-bound this variance by the second moments using
\begin{equation}
    \Delta^2_{\ket{\psi}} \hat G(\phi)
    =
    \expval{\hat G(\phi)^2}_{\psi}
    - \expval{\hat G(\phi)}_{\psi}^2
    \le
    \expval{\hat G(\phi)^2}_{\psi}.
    \label{eq:var-le-second-moment}
\end{equation}

The phase-averaged state $\hat\rho_{\boldsymbol{\alpha}}$ in \eqref{eq:phase-avg-state} is a convex mixture of pure states:
\begin{equation}
    \hat\rho_{\boldsymbol{\alpha}}^\text{pure}
    =
    \int_0^{2\pi} \frac{d\phi}{2\pi}\,
    \ket{\psi_{\boldsymbol{\alpha}}^\phi}\bra{\psi_{\boldsymbol{\alpha}}^\phi}.
\end{equation}
The QFIM is a convex function of the state, in the sense that for any family of states $\{\hat\sigma^{(\ell)}\}$ and convex weights $\{p_\ell\}$,
\begin{equation}
    F_Q^{(\boldsymbol{\alpha})}\left(\sum_\ell p_\ell \hat\sigma^{(\ell)}\right)
    \preceq
    \sum_\ell p_\ell F_Q^{(\boldsymbol{\alpha})}(\hat\sigma^{(\ell)}).
\end{equation}
Applying this to the continuous mixture over $\phi$, we obtain the matrix inequality
\begin{equation}
    F_Q^{(\boldsymbol{\alpha})}(\hat\rho_{\boldsymbol{\alpha}})
    \preceq
    \int_0^{2\pi} \frac{d\phi}{2\pi}\,
    F_Q^{(\boldsymbol{\alpha})}\left(\ket{\psi^\phi_{\boldsymbol{\alpha}}}\right).
\end{equation}
Multiplying by $\mathbf{1}^T$ from the left and $\mathbf{1}$ from the right,
\begin{equation}
    \mathbf{1}^T F_Q^{(\boldsymbol{\alpha})}(\hat\rho_{\boldsymbol{\alpha}})\mathbf{1}
    \le
    \int_0^{2\pi} \frac{d\phi}{2\pi}\,
    \mathbf{1}^T
    F_Q^{(\boldsymbol{\alpha})}\left(\ket{\psi^\phi_{\boldsymbol{\alpha}}}\right)
    \mathbf{1}.
\end{equation}
Using \eqref{eq:FQ-alpha-as-quadratic-form} and \eqref{eq:scalar-QFI-pure},
\begin{equation}
    F_Q^{(\alpha)}(\hat\rho_{\alpha})
    =
    \mathbf{1}^T F_Q^{(\boldsymbol{\alpha})}(\hat\rho_{\boldsymbol{\alpha}})\mathbf{1}
    \le
    \int_0^{2\pi}\frac{d\phi}{2\pi}\,
    4\,\Delta^2_{\ket{\psi}}\hat G(\phi).
\end{equation}
Using \eqref{eq:var-le-second-moment},
\begin{equation}
    F_Q^{(\alpha)}(\hat\rho_{\alpha})
    \le
    4\int_0^{2\pi}\frac{d\phi}{2\pi}\,
    \expval{\hat G(\phi)^2}_{\psi}.
    \label{eq:QFI-bound-second-moment}
\end{equation}
We now compute the phase average of the second moment explicitly.
A straightforward calculation gives
\begin{align}
    \hat A_i(\phi)\hat A_j(\phi)
    &=
    -e^{i2\phi}\hat a_i^\dagger \hat a_j^\dagger
    -e^{-i2\phi}\hat a_i \hat a_j
    +\hat a_i^\dagger \hat a_j
    +\hat a_i \hat a_j^\dagger.
\end{align}
Therefore
\begin{align}
    \hat G(\phi)^2
    &= \sum_{i,j=1}^M \hat A_i(\phi)\hat A_j(\phi)
\nonumber\\
    &=
    -e^{i2\phi}\left(\sum_{i=1}^M \hat a_i^\dagger\right)^2
    -e^{-i2\phi}\left(\sum_{i=1}^M \hat a_i\right)^2
\nonumber\\
    &\quad
    +\sum_{i,j=1}^M \left(\hat a_i^\dagger \hat a_j + \hat a_i \hat a_j^\dagger\right).
\end{align}
Taking the expectation value in $\ket{\psi}$ and inserting into \eqref{eq:QFI-bound-second-moment}, we obtain
\begin{align}
    F_Q^{(\alpha)}(\hat\rho_{\alpha})
    &\le
    4\int_0^{2\pi}\frac{d\phi}{2\pi}\,
    \Bigl[
      -e^{i2\phi}\expval{\Bigl(\sum_i \hat a_i^\dagger\Bigr)^2}
      -e^{-i2\phi}\expval{\Bigl(\sum_i \hat a_i\Bigr)^2}
\nonumber\\
    &\qquad\qquad\qquad\qquad
      +\expval{\sum_{i,j}(\hat a_i^\dagger \hat a_j + \hat a_i \hat a_j^\dagger)}
    \Bigr].
\end{align}
The terms proportional to $e^{\pm i2\phi}$ vanish under the integral, leaving
\begin{equation}
    F_Q^{(\alpha)}(\hat\rho_{\alpha})
    \le
    4\expval{\sum_{i,j=1}^M(\hat a_i^\dagger \hat a_j + \hat a_i \hat a_j^\dagger)}.
    \label{eq:FQ-bound-raw-matrix}
\end{equation}

We decompose the sum over $i,j$ into diagonal ($i=j$) and off-diagonal ($i\neq j$) parts:
\begin{align}
    \sum_{i,j}(\hat a_i^\dagger \hat a_j + \hat a_i \hat a_j^\dagger)
    &=
    \sum_{i=1}^M (\hat a_i^\dagger \hat a_i + \hat a_i \hat a_i^\dagger)
    +\sum_{i\neq j}(\hat a_i^\dagger \hat a_j + \hat a_i \hat a_j^\dagger)
\nonumber\\
    &=
    \sum_{i=1}^M (2\hat a_i^\dagger \hat a_i + \mathbbm{1})
    +\sum_{i\neq j}(\hat a_i^\dagger \hat a_j + \hat a_i \hat a_j^\dagger),
\end{align}
since $\hat a_i \hat a_i^\dagger = \hat a_i^\dagger \hat a_i + \mathbbm{1}$.
Let $\hat n_i := \hat a_i^\dagger \hat a_i$. Taking expectation values,
\begin{align}
    F_Q^{(\alpha)}(\hat\rho_{\alpha})
    &\le 4\sum_{i=1}^M\left(2\expval{\hat n_i} + 1\right)
\nonumber\\
    &\quad
      +4\sum_{i\neq j}\expval{\hat a_i^\dagger \hat a_j + \hat a_i \hat a_j^\dagger}.
    \label{eq:FQ-bound-diag-offdiag}
\end{align}

For $i\neq j$, apply the Cauchy--Schwarz inequality to $\hat X = \hat a_i$ and $\hat Y = \hat a_j$:
\begin{equation}
    \abs{\expval{\hat a_i^\dagger \hat a_j}}^2
    \le
    \expval{\hat a_i^\dagger \hat a_i}
    \expval{\hat a_j^\dagger \hat a_j}
    =
    \expval{\hat n_i}\expval{\hat n_j} \qquad\Rightarrow\qquad     \abs{\expval{\hat a_i^\dagger \hat a_j}}
    \le
    \sqrt{\expval{\hat n_i}\expval{\hat n_j}}.
\end{equation}
Moreover, note that
\begin{equation}
    \expval{\hat a_i^\dagger \hat a_j + \hat a_i \hat a_j^\dagger}
    =
    2\mathrm{Re}\expval{\hat a_i^\dagger \hat a_j}
    \le
    2\abs{\expval{\hat a_i^\dagger \hat a_j}}
    \le 2\sqrt{\expval{\hat n_i}\expval{\hat n_j}}.
\end{equation}
Summation over $i\neq j$ yields
\begin{equation}
    \sum_{i\neq j}\expval{\hat a_i^\dagger \hat a_j + \hat a_i \hat a_j^\dagger}
    \le 2\sum_{i\neq j}\sqrt{\expval{\hat n_i}\expval{\hat n_j}}.
    \label{eq:offdiag-CS}
\end{equation}
Define
\begin{equation}
    S := \sum_{i=1}^M \sqrt{\expval{\hat n_i}} \qquad\Rightarrow\qquad     S^2
    = \sum_{i=1}^M \expval{\hat n_i}
      + 2\sum_{i\neq j}\sqrt{\expval{\hat n_i}\expval{\hat n_j}}  \quad\Rightarrow\quad \sum_{i\neq j}\sqrt{\expval{\hat n_i}\expval{\hat n_j}}
    = \frac{1}{2}\left(S^2 - \sum_{i=1}^M \expval{\hat n_i}\right) .
\end{equation}
On the other hand, we can apply the Cauchy--Schwarz to the vector
$(\sqrt{\expval{\hat n_1}},\dots,\sqrt{\expval{\hat n_M}})$, to obtain
\begin{equation}
    S^2
    =
    \left(\sum_{i=1}^M \sqrt{\expval{\hat n_i}}\right)^2
    \le
    M\sum_{i=1}^M \expval{\hat n_i}.
\end{equation}
Hence
\begin{align}
    \sum_{i\neq j}\sqrt{\expval{\hat n_i}\expval{\hat n_j}}
    &\le
    \frac{1}{2}\left[
      M\sum_{i=1}^M \expval{\hat n_i}
      - \sum_{i=1}^M \expval{\hat n_i}
    \right]
\nonumber\\
    &= \frac{M-1}{2}\sum_{i=1}^M \expval{\hat n_i}.
\end{align}
Plugging this into \eqref{eq:offdiag-CS},
\begin{equation}
    \sum_{i\neq j}\expval{\hat a_i^\dagger \hat a_j + \hat a_i \hat a_j^\dagger}
    \le
    2\cdot\frac{M-1}{2}\sum_{i=1}^M \expval{\hat n_i}
    =
    (M-1)\sum_{i=1}^M \expval{\hat n_i}.
\end{equation}

Substituting into \eqref{eq:FQ-bound-diag-offdiag}, we obtain
\begin{align}
    F_Q^{(\alpha)}(\hat\rho_{\alpha})
    &\le
    4\sum_{i=1}^M\left(2\expval{\hat n_i} + 1\right)
    +4(M-1)\sum_{i=1}^M \expval{\hat n_i}
\nonumber\\
    &=
    4M
    + \left(8 + 4(M-1)\right)\sum_{i=1}^M \expval{\hat n_i}
\nonumber\\
    &=
    4M
    + 4(M+1)\sum_{i=1}^M \expval{\hat n_i}.
\end{align}
A slightly looser but symmetric bound is obtained by noticing $M+1\le 2M$ for $M\ge 1$, giving
\begin{equation}
    F_Q^{(\alpha)}(\hat\rho_{\alpha})
    \le
    4M
    + 8M\sum_{i=1}^M \expval{\hat n_i}.
\end{equation}
Equivalently, in terms of the total excitation number
\begin{equation}
    \hat N
    :=
    \sum_{i=1}^M \hat n_i,
    \qquad
    \expval{\hat N}
    = \sum_{i=1}^M \expval{\hat n_i},
\end{equation}
we can write
\begin{equation}
    F_Q^{(\alpha)}(\hat\rho_{\alpha})
    \le
    4M + 8M\expval{\hat N}
    = 4M\bigl(2\expval{\hat N}+1\bigr).
    \label{eq:FQ-bound-final}
\end{equation}

For classical probe states, i.e., mixtures of product coherent states,
\begin{equation}
    \hat\rho_{\text{cl}}
    =
    \int d\beta\,P(\beta)\bigotimes_{i=1}^M \ket{\beta_i}\bra{\beta_i},
    \qquad P(\beta)\ge 0,
\end{equation}
the main text shows that the QFI for $\alpha$ satisfies
\begin{equation}
    F_Q^{(\alpha)}(\hat\rho_{\text{cl}}) \le 4M.
\end{equation}
This value
\begin{equation}
    F_Q^{\text{SQL}} := 4M
\end{equation}
is then identified as the \emph{standard quantum limit} (SQL) for the phase-insensitive distributed force sensing task.

Dividing \eqref{eq:FQ-bound-final} by $F_Q^{\text{SQL}}$ yields
\begin{equation}
    \frac{F_Q^{(\alpha)}(\hat\rho_{\alpha})}{F_Q^{\text{SQL}}}
    =
    \frac{F_Q^{(\alpha)}(\hat\rho_{\alpha})}{4M}
    \le
    2\expval{\hat N}+1.
    \label{eq:eq10}
\end{equation}
Equation \eqref{eq:eq10} is exactly Eq.~(10) in the main text, now derived explicitly via the QFI \emph{matrix} formalism. The key observation is that the single-parameter QFI for the global amplitude $\alpha$ is the quadratic form
\begin{equation}
    F_Q^{(\alpha)} = \mathbf{1}^T F_Q^{(\boldsymbol{\alpha})}\mathbf{1},
\end{equation}
and the energy constraint on the total excitation number $\expval{\hat N}$ imposes the bound \eqref{eq:FQ-bound-final} on this quadratic form.

\end{document}